\documentclass[aip, jcp, amsmath,amssymb, reprint, floatfix,citeautoscript]{revtex4-1}

\usepackage{graphicx}
\usepackage{dcolumn}
\usepackage{bm}

\usepackage{braket}
\usepackage[usenames,dvipsnames]{color}

\usepackage{ragged2e}
\usepackage[labelsep=period, format=plain, justification=justified, font=footnotesize]{caption}
\usepackage{times}
\usepackage{txfonts}
\usepackage{relsize}
\usepackage{multirow}
\usepackage{tabularx,booktabs}
\newcolumntype{Y}{>{\centering\arraybackslash}X}

\DeclareCaptionJustification{myjust}{\justifying}
\newcommand{\RNum}[1]{\uppercase\expandafter{\romannumeral #1\relax}}

\begin{document}

\title{Accurate and efficient DFT-based diabatization for hole and electron transfer using absolutely localized molecular orbitals}

\author{Yuezhi Mao}
\author{Andr\'{e}s Montoya-Castillo}
\author{Thomas E. Markland}
\email{tmarkland@stanford.edu}
\affiliation{Department of Chemistry, Stanford University, Stanford, California, 94305, USA}

\date{\today}

\begin{abstract}
Diabatic states and the couplings between them are important for quantifying, elucidating, and predicting the rates and mechanisms of many chemical and biochemical processes. Here, we propose and investigate approaches to accurately compute diabatic couplings from density functional theory (DFT) using absolutely localized molecular orbitals (ALMOs). ALMOs provide an appealing approach to generate variationally optimized diabatic states and obtain their associated forces that allows for the relaxation of the donor and acceptor orbitals in a way that is internally consistent in how the method treats both the donor and acceptor states. Here, we show that one can obtain more accurate electronic couplings between ALMO-based diabats by employing the symmetrized transition density matrix to evaluate the exchange-correlation contribution. We demonstrate that this approach yields accurate results in comparison to other commonly used DFT-based diabatization methods across a wide array of electron and hole transfer processes occurring in systems ranging from conjugated organic molecules, such as thiophene and pentacene, to DNA base pairs. We also show that this approach yields accurate diabatic couplings even when combined with lower tiers of the DFT hierarchy, opening the door to combining it with quantum dynamics approaches to provide an ab initio treatment of nonadiabatic processes in the condensed phase.
\end{abstract}

{\maketitle}
\normalsize

\section{Introduction}

Electron transfer (ET) and hole transfer (HT) are fundamental steps in many chemical and biochemical processes, ranging from charge and energy transport in photovoltaic materials to electrocatalysis and enzyme-catalyzed reactions. A convenient way to describe ET and HT reactions relies on employing charge-localized donor and acceptor states corresponding to reactants and products. Importantly, these chemically intuitive diabatic states retain their character along the reaction coordinate, whereas adiabatic states, which are the natural output of electronic structure calculations, do not. Diabatic states thus form the basis of widely used theories of reaction rates, including Marcus-Hush theory of electron transfer\cite{Marcus1956, Hush1961, Marcus1993}, excitation energy transfer theory\cite{Forster1948}, and Marcus theory-inspired approaches to proton\cite{Borgis1989,Borgis1993} and proton-coupled electron transfer\cite{Soudackov2000,Hammes-Schiffer2001}. These rate theories have proven critical in elucidating dependence of rates and mechanisms on the microscopic parameters of diverse chemical and biochemical reactions\cite{Adams2003, Migliore2014}, and continue to serve as the major workhorses for the understanding and rational design of new chemical systems\cite{Wang2010, Hammes-Schiffer2018}. However, as new experimental techniques yield increasingly detailed time-resolved measurements\cite{Jonas2003, Cheng2009, Kowalewski2017, Carpenter2016}, the need for quantum dynamical information that goes beyond rates becomes more apparent. In such cases, diabatic potential energy surfaces (PESs) and their associated forces offer the means to connect an ab initio description of chemical systems with quantum dynamics approaches for the simulation of nonadiabatic processes, while circumventing the difficulties associated with adiabatic states such as those arising from the diverging derivative coupling when adiabatic states approach each other. However, since electronic structure theory relies on diagonalizing the electronic part of the Hamiltonian, which naturally yields adiabatic states, one needs to resort to diabatization schemes to obtain diabatic states from ab initio calculations.

One class of diabatization schemes aims to construct diabatic states by generating a unitary transformation from adiabatic states obtained from an electronic structure calculation.\cite{Baer1980,Pacher1988,Pacher1991,Ruedenberg1993,Nakamura2001,Cave1996,Cave1997,Voityuk2002,Hsu2008,Hsu2009,You2010,Subotnik2008,Subotnik2009,Subotnik2010,Subotnik2015} Although one is almost never able to build strictly diabatic states, i.e., those with zero derivative couplings, from such transformations,\cite{Mead1982} one can still obtain approximate diabatic states with correct charge-localization character. For example, the adiabatic-to-diabatic transformation can be constructed by defining the diabatic states as eigenstates of a relevant molecular property operator,\cite{Cave1996,Cave1997,Voityuk2002,Hsu2008,Hsu2009,You2010} such as the dipole operator along the charge-transfer direction in the generalized Mulliken-Hush (GMH) method,\cite{Cave1996,Cave1997} or as states that maximize objective functions analogous to those employed in Boys\cite{Foster1960} or Edmiston-Ruedenberg\cite{Edmiston1963} orbital localization schemes. \cite{Subotnik2008,Subotnik2009,Subotnik2010,Subotnik2015} The accuracy of these approaches is particularly sensitive to the quality of the excited-state method employed. As ET/HT complexes are intrinsically of a multireference character owing to the degeneracy or near-degeneracy of the donor and acceptor states, high-level electronic structure methods that account for static electron correlation are usually required to obtain a reliable set of adiabats, limiting the use of these diabatization schemes to small molecular systems. In addition, since the adiabatic-to-diabatic transformation matrix changes with the nuclear positions, it also needs to be taken into account in the force calculation, making it theoretically challenging and computationally expensive to compute forces for diabatic states transformed from an adiabatic basis.

As Kohn-Sham density functional theory (KS-DFT)\cite{Kohn1965,Kohn1996} can efficiently incorporate dynamical electron correlation at mean-field cost, many approaches have been developed to obtain diabatic states and couplings from DFT calculations. These methods rely on partitioning either the electron density or the orbital space of the full system. Some of these methods focus primarily on extracting the electronic coupling between diabatic states, whereas others generate variationally optimized many-electron diabatic wavefunctions in addition to their couplings, thus providing convenient access to forces.

One method belonging to the former category, the projection-operator diabatization (POD) approach,\cite{Kondov2007,Futera2017} utilizes a partition of atomic orbital (AO) basis functions into donor ($D$) and acceptor ($A$) groups. In this method, one starts from a converged KS-DFT calculation of the closed-shell ground state, diagonalizes the $DD$ and $AA$ blocks of the KS Fock matrix to generate the diabatic molecular orbitals (MOs), and then transforms the $DA$ block into this new MO basis. The $DA$ block, when expressed in this new MO basis, contains the couplings between pairs of single-particle orbitals, which represents the diabatic coupling. Since the diabatic states are approximated by single-particle MOs, one is unable to construct many-electron diabatic PESs with this scheme.

The fragment-orbital DFT (FODFT) method\cite{Senthilkumar2003,Oberhofer2012,Schober2016} falls in the same category as POD, but does not require optimization of the orbitals of the full system. Instead, in FODFT one separately generates the optimized orbitals for the donor and acceptor fragments. The diabatic coupling is then approximated as the coupling through the KS Fock operator, $\hat{f}_{\text{KS}}$, between the pair of fragment orbitals involved in the charge transfer process. Since there is no unique prescription for the preparation of fragment orbitals and the construction of $\hat{f}_{\text{KS}}$, three flavors of FODFT have been proposed (see Sec.~\ref{sec:comput_detail} as well as Ref.~\citenum{Schober2016}). In the most accurate variant of FODFT\cite{Schober2016} only one of the charge-localized states (the donor or acceptor) is used to construct both the fragment orbitals and $\hat{f}_{\text{KS}}$. This creates the problem that, for systems without explicit symmetry, different electronic couplings are obtained depending on the choice of charge-localized state used, i.e., 
\begin{equation}
    H_{ab} = \braket{\psi_a | \hat{H} | \psi_b} \neq \braket{\psi_b | \hat{H} | \psi_a} = H_{ba},
\end{equation}
where the subscripts $a$ and $b$ label the two diabatic states. Also, since the donor and acceptor orbitals are not allowed to relax when they are brought together, FODFT will become less accurate for systems where the donor and acceptor interact strongly.

To obtain forces associated with those diabatic PESs, it is advantageous to work with variationally optimized many-electron wavefunctions, which is not the case for either the POD or FODFT schemes. Access to forces allows one to perform geometry optimization and molecular dynamics on the diabatic surfaces to calculate Marcus parameters such as the reorganization energy and the driving force for the charge transfer reaction.\cite{Wu2006c,VanVoorhis2010,Oberhofer2009,Kowalczyk2011,Rezac2012,Holmberg2017} Two approaches which produce variationally optimized wavefunctions are constrained DFT (CDFT)\cite{Dederichs1984,Wu2005,Wu2006b,Kaduk2011} and schemes based on absolutely localized molecular orbitals (ALMO)\cite{Khaliullin2006} / block-localized wavefunctions (BLW).\cite{Mo1998, Mo2007} 
CDFT produces diabatic states by imposing constraints on the real-space electron density in KS-DFT calculations, which is achieved by introducing Lagrangian multipliers into the energy functional.\cite{Wu2005,Wu2006b} The electronic coupling between two CDFT diabatic states can then be computed by approximating each state as an eigenfunction of the electronic Hamiltonian with the corresponding constraining potential.\cite{Wu2006} Previous benchmarks have demonstrated that CDFT is able to give accurate $H_{ab}$ values for ET and HT, although its performance depends sensitively on the exchange-correlation (XC) functional employed.\cite{Kubas2014,Kubas2015} Moreover, CDFT has been shown to fail to predict the correct long-range behavior in systems that have degenerate or near-degenerate frontier orbitals as the constraint on electron density is insufficient to exclude the unphysical scenario where fractions of multiple electrons are transferred\cite{Mavros2015}.

To circumvent the unphysical behavior of transferring parts of multiple electrons one can impose constraints in the orbital space instead. ALMOs provide a solution to this unphysical behavior by expanding the MOs on each fragment using the atomic orbital (AO) basis functions on that fragment alone.\cite{Khaliullin2006} The ALMO (also known as BLW) states can then be variationally optimized\cite{Stoll1980,Gianinetti1996} subject to the constraint that the charges are fragment-localized according to the Mulliken definition\cite{Mulliken1955}. While both ALMO- and FODFT-based diabatic states share the property that the MOs are ``absolutely localized" on fragments, in the former approach states are variationally optimized with respect to orbital rotations, which allows the donor and acceptor orbitals to relax when brought together. Once constructed, one can apply the multistate DFT (MSDFT) approach\cite{Cembran2009} to approximate the electronic coupling between two ALMO diabatic states.\cite{Ren2016,Guo2018} However, this scheme systematically overestimates the magnitude of diabatic coupling, as has been previously demonstrated in the case of HT \cite{Guo2018} and as we show in Sec.~\ref{sec:results} for both ET and HT, which motivates the need to develop a more accurate scheme.

Here, we propose an improved scheme to evaluate the electronic couplings between ALMO-based diabatic states. We show that by using the symmetrized transition density between two diabatic states, one can account for the XC contribution to the off-diagonal elements of the diabatic Hamiltonian more accurately. We demonstrate the performance of this new approach by comparing it to the POD, FODFT, CDFT, and previous MSDFT approaches for the diabatic couplings in a wide range of ET and HT systems. This improved approach yields accurate diabatic couplings and provides access to variationally optimized states and forces. We show that this accuracy holds even when combined with lower-tier XC functionals, thus providing a computationally efficient approach to obtain accurate diabatic couplings. 

\section{Methods} \label{sec:method}

In this section, we first summarize the procedure needed to construct variationally optimized diabatic states using ALMOs. We then introduce two approaches to calculate the electronic coupling between these ALMO-based diabatic states. In the first approach, ALMO(FODFT), we combine the FODFT approach with variationally optimized ALMO-based diabats. We then detail the ALMO(MSDFT) approach and suggest an improved procedure for calculating diabatic couplings from it, which we denote as ALMO(MSDFT2).  

\subsection{Charge-localized diabatic states from ALMOs}

In this work we construct charge-localized diabatic states using ALMOs. Here we illustrate this approach with the example of HT in a donor-acceptor system where the neutral state of the donor has $n_D$ electrons and that of the acceptor has $n_A$ electrons with the full system having $N = n_D + n_A - 1$ electrons in total. With a partition of the supersystem into the donor ($D$) and acceptor ($A$) fragments, the two diabats are of the following forms:
\begin{subequations}
\begin{align}
    \ket{\psi_a} &= \frac{1}{\sqrt{(N-1)!}} \mathrm{det} \left\lbrace \phi^{(a)}_{D1}, \phi^{(a)}_{D2}, \dots, \phi^{(a)}_{Dn_{D}\mathrm{-}1}  \phi^{(a)}_{A1}, \phi^{(a)}_{A2}, \dots, \phi^{(a)}_{An_{A}} \right \rbrace \label{eq:almo_donor}\\
    \ket{\psi_b} &= \frac{1}{\sqrt{(N-1)!}} \mathrm{det}  \left\lbrace \phi^{(b)}_{D1}, \phi^{(b)}_{D2}, \dots, \phi^{(b)}_{Dn_{D}}  \phi^{(b)}_{A1}, \phi^{(b)}_{A2}, \dots, \phi^{(b)}_{An_{A}\mathrm{-}1} \right \rbrace \label{eq:almo_acceptor}
\end{align}
\end{subequations}
where $\ket{\psi_a}$ and $\ket{\psi_b}$ correspond to the reactant ($D^{+}A$) and product ($DA^{+}$) diabats, respectively, and ``det" denotes the Slater determinants. Each MO in the determinant is ``absolutely localized" on either the donor or the acceptor fragment as indicated by the fragment label in its subscript, i.e., the MOs on a particular fragment are expanded by AO basis functions assigned to that fragment. The superscripts, $(a)$ and $(b)$, indicate that the ALMOs are optimized within each individual diabatic configuration such that the donor and acceptor sets of orbitals in $\psi_{a}$ and $\psi_{b}$ differ from one another.

\begin{figure}
    \centering
    \includegraphics[width=0.48\textwidth]{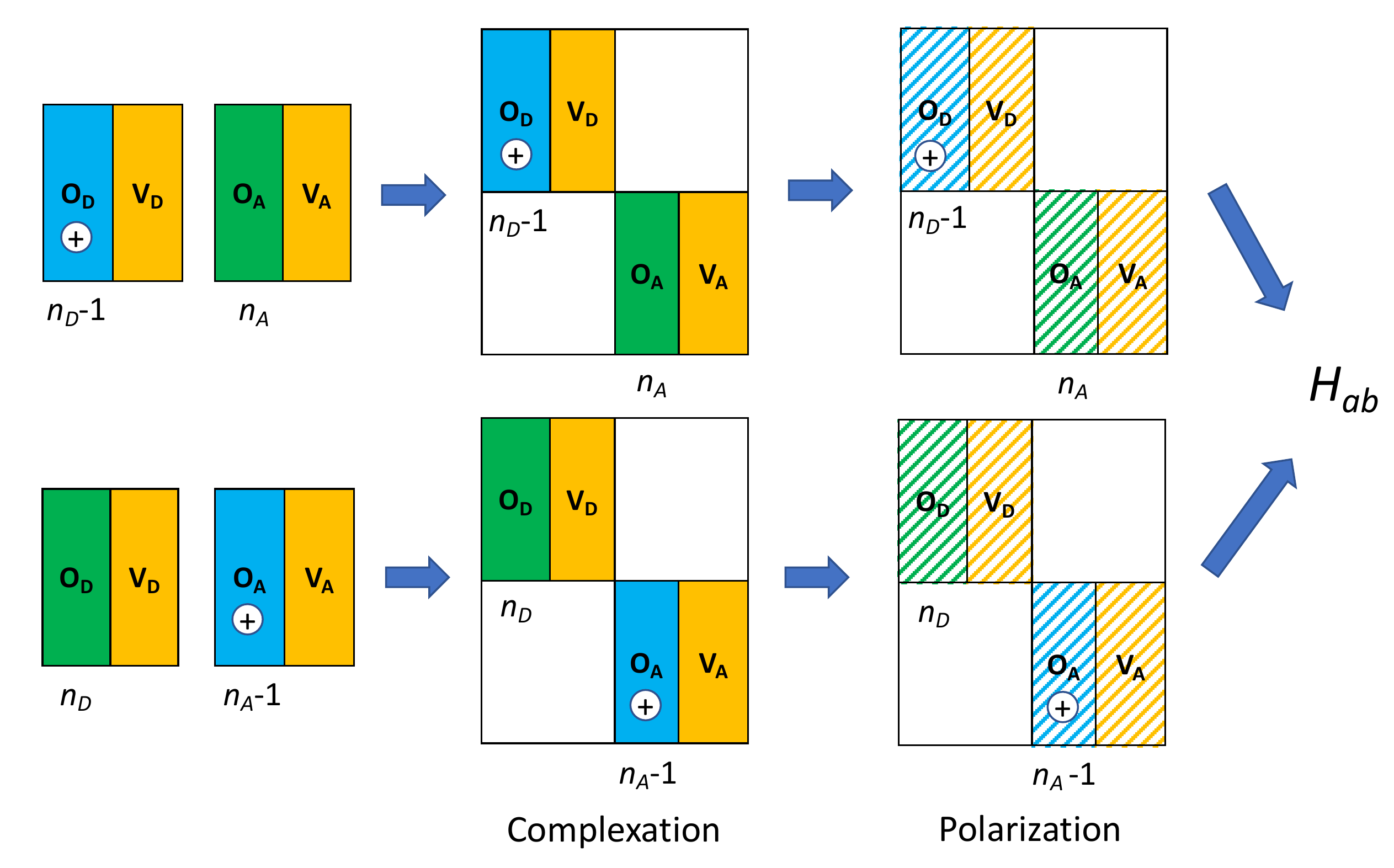}
    \caption{The ALMO(MSDFT) procedure, including: (i) fragment SCF calculations for $D^+$, $A$ and $D$, $A^+$; (ii) construction of two diabats $\ket{D^+A}$ and $\ket{DA^+}$; (iii) variational relaxation of two ALMO states; (iv) evaluation of $H_{ab}$ using the MSDFT scheme. }
    \label{fig:almo_msdft}
\end{figure}

With well-separated donor and acceptor fragments, one can compute the ALMOs of the interacting donor-acceptor system in a ``bottom-up" fashion, i.e., by starting from the orbitals obtained from self-consistent field (SCF) calculations of the isolated fragments that comprise each diabat ($D^{+}$ and $A$ fragments for one diabat and $D$ and $A^{+}$ for the other in the case of HT). One can then generate the initial guess for each of the diabats by assembling the fragment orbitals into one single antisymmetrized product, which corresponds to constructing the MO coefficients for the full system by concatenating the fragment MO coefficients (see the ``complexation" step in Figs.~\ref{fig:almo_fodft} and \ref{fig:almo_msdft}). This state is referred to as the frozen wavefunction in ALMO-based energy decomposition analysis.\cite{Khaliullin2007, Horn2016c} For example, to build the $D^{+}A$ diabatic wavefunction ($\ket{\psi_a}$), one concatenates the fragment orbitals formed on the isolated $D^{+}$ and $A$ fragments. Its associated one-particle density matrix (1PDM) is
\begin{equation}
    \mathbf{P}^{(a)} = \mathbf{C}^{(a)}_{\text{o}} (\boldsymbol{\sigma}^{(a)}_{\text{oo}})^{-1} (\mathbf{C}^{(a)}_{\text{o}})^{T}, \label{eq:almo_den}
\end{equation}
where $\mathbf{C}^{(a)}_{\text{o}}$ refers to the MO coefficients for the occupied orbitals in $\ket{\psi_a}$ and $\boldsymbol{\sigma}^{(a)}_{\text{oo}}$ denotes the overlap metric for them. The overlap metric can be obtained by transforming the AO overlap matrix ($\mathbf{S}$) into the basis formed by occupied fragment orbitals:
\begin{equation}
    \boldsymbol{\sigma}^{(a)}_{\text{oo}} = (\mathbf{C}^{(a)}_{\text{o}})^{T} \mathbf{S} \mathbf{C}^{(a)}_{\text{o}}.
\end{equation}
It is straightforward to show that a 1PDM constructed through Eq.~(\ref{eq:almo_den}) satisfies
\begin{subequations}
\begin{align}
    \mathrm{Tr}[\mathbf{P^{(a)}S}]_{DD} &= \mathrm{Tr}[\mathbf{I}^{(a)}]_{D} = n_{D}-1, \\
    \mathrm{Tr}[\mathbf{P^{(a)}S}]_{AA} &= \mathrm{Tr}[\mathbf{I}^{(a)}]_{A} = n_{A}.
\end{align}
\end{subequations}
Correspondingly for $\ket{\psi_b}$ constructed from the $D$ and $A^{+}$ fragments
\begin{subequations}
\begin{align}
    \mathrm{Tr}[\mathbf{P^{(b)}S}]_{DD} &= \mathrm{Tr}[\mathbf{I}^{(b)}]_{D} = n_{D}, \\
    \mathrm{Tr}[\mathbf{P^{(b)}S}]_{AA} &= \mathrm{Tr}[\mathbf{I}^{(b)}]_{A} = n_{A}-1.
\end{align}
\end{subequations}
Hence, under the Mulliken definition,\cite{Mulliken1955} the charge populations on the donor and acceptor fragments are unchanged from their values in the isolated state.

The energy of an ALMO-based diabatic state can then be evaluated from its 1PDM:
\begin{equation}
    E^{\text{KS}}[\mathbf{P}] = V_{\text{nn}} + \mathbf{P}\cdot\mathbf{h} + \frac{1}{2} \mathbf{P}\cdot\mathrm{\RNum{2}}\cdot\mathbf{P} + E_{\text{xc}}[\mathbf{P}], \label{eq:ks_energy}
\end{equation}
where $\mathbf{h}$ is the core-Hamiltonian, $\mathrm{\RNum{2}}$ represents the two-electron integrals, $E_{\text{xc}}$ is the exchange-correlation energy functional, and $V_{\text{nn}}$  refers to the nuclear repulsion energy. Starting from the frozen state, one can variationally optimize a diabatic wavefunction by minimizing $E[\mathbf{P}]$ with respect to the on-fragment occupied-virtual orbital mixings, which can be achieved by solving locally projected SCF equations\cite{Stoll1980,Gianinetti1996} or employing gradient-based optimization methods.\cite{Horn2015} By relaxing the ALMOs for the $D^+A$ and $DA^+$ configurations separately, one takes into account the mutual polarization between the donor and acceptor fragments and obtains the two diabatic wavefunctions represented in Eqs.~(\ref{eq:almo_donor}) and (\ref{eq:almo_acceptor}).

Finally, we note that it is also possible to construct ALMO states using a ``top-down" procedure, i.e., from fully relaxed SCF solutions for the supersystem.\cite{Mo2010, Zhang2016, Mao2018} This is necessary for systems without a clear partition between the donor and acceptor fragments, such as intramolecular ET/HT through a bridging moiety. In this work, we focus on intermolecular ET/HT systems and employ only ``bottom-up" approach. However, the schemes to compute the diabatic couplings introduced below are equally applicable when employing the ``top-down" procedure.

\subsection{ALMO(FODFT) approach} \label{subsec:almo_fodft}

The most recent variant of FODFT\cite{Schober2016} utilizes fragment orbitals evaluated on the charged donor ($D^+$) and the neutral acceptor ($A$). The supersystem wavefunction constructed from these orbitals is thus identical to the frozen state (initial guess) for the ALMO-based diabat shown in Eq.~(\ref{eq:almo_donor}). Here we propose a modification to the original FODFT scheme: instead of using unrelaxed fragment orbitals, we perform FODFT calculations for diabatic couplings on top of variationally optimized ALMO states so that the orbitals on the donor and acceptor fragments are relaxed in the presence of each other. We call this scheme ALMO(FODFT) in the following discussion.

To calculate the diabatic coupling ($H_{ab}$) with FODFT, one first orthogonalizes the fragment orbitals calculated from $D^+$ and $A$ using L\"owdin's symmetric orthogonalization scheme.\cite{Lowdin1950} Note that the lowest empty orbital, which corresponds to the hole, $\phi_{Dn_{D}}^{(a)}$, also needs to be made orthogonal to all the occupied orbitals. Here, we achieve this by projecting out the space spanned by the occupied MOs from the hole orbital and then renormalizing it. Using orthogonalized orbitals, the reactant state ($D^+A$) can be written as
\begin{equation}
    \ket{\bar{\psi}_a} = \frac{1}{\sqrt{(N-1)!}} \mathrm{det} \left\lbrace \bar{\phi}^{(a)}_{D1}, \bar{\phi}^{(a)}_{D2}, \dots, \bar{\phi}^{(a)}_{Dn_{D}\mathrm{-}1}  \bar{\phi}^{(a)}_{A1}, \bar{\phi}^{(a)}_{A2}, \dots, \bar{\phi}^{(a)}_{An_{A}} \right \rbrace, \label{eq:fodft_donor}
\end{equation}
where the bar denotes L\"owdin-orthogonalized fragment orbitals. The product diabat ($DA^+$) is then approximated by moving the electron from the $n_A$th orbital of $A$ to the $n_D$th orbital of $D$:
\begin{equation}
    \ket{\bar{\psi}_b} = \frac{1}{\sqrt{(N-1)!}} \mathrm{det}  \left\lbrace \bar{\phi}^{(a)}_{D1}, \bar{\phi}^{(a)}_{D2}, \dots, \bar{\phi}^{(a)}_{Dn_{D}}  \bar{\phi}^{(a)}_{A1}, \bar{\phi}^{(a)}_{A2}, \dots, \bar{\phi}^{(a)}_{An_{A}\mathrm{-}1} \right \rbrace. \label{eq:fodft_acceptor}
\end{equation}
The diabatic coupling between these two states can then be approximated by coupling the frontier orbitals ($\bar{\phi}_{Dn_{D}}^{(a)}$ and $\bar{\phi}_{An_{A}}^{(a)}$) through the KS Fock operator ($\hat{f}_{\text{KS}}$),
\begin{equation}
    H_{ab} = \braket{\bar{\psi}_a | \hat{H} | \bar{\psi}_b} \approx \braket{\bar{\phi}_{An_{A}}^{(a)} | \hat{f}_{\text{KS}} | \bar{\phi}_{Dn_{D}}^{(a)}}. \label{eq:fodft_coupling}
\end{equation}

\begin{figure}
    \centering
    \includegraphics[width=0.45\textwidth]{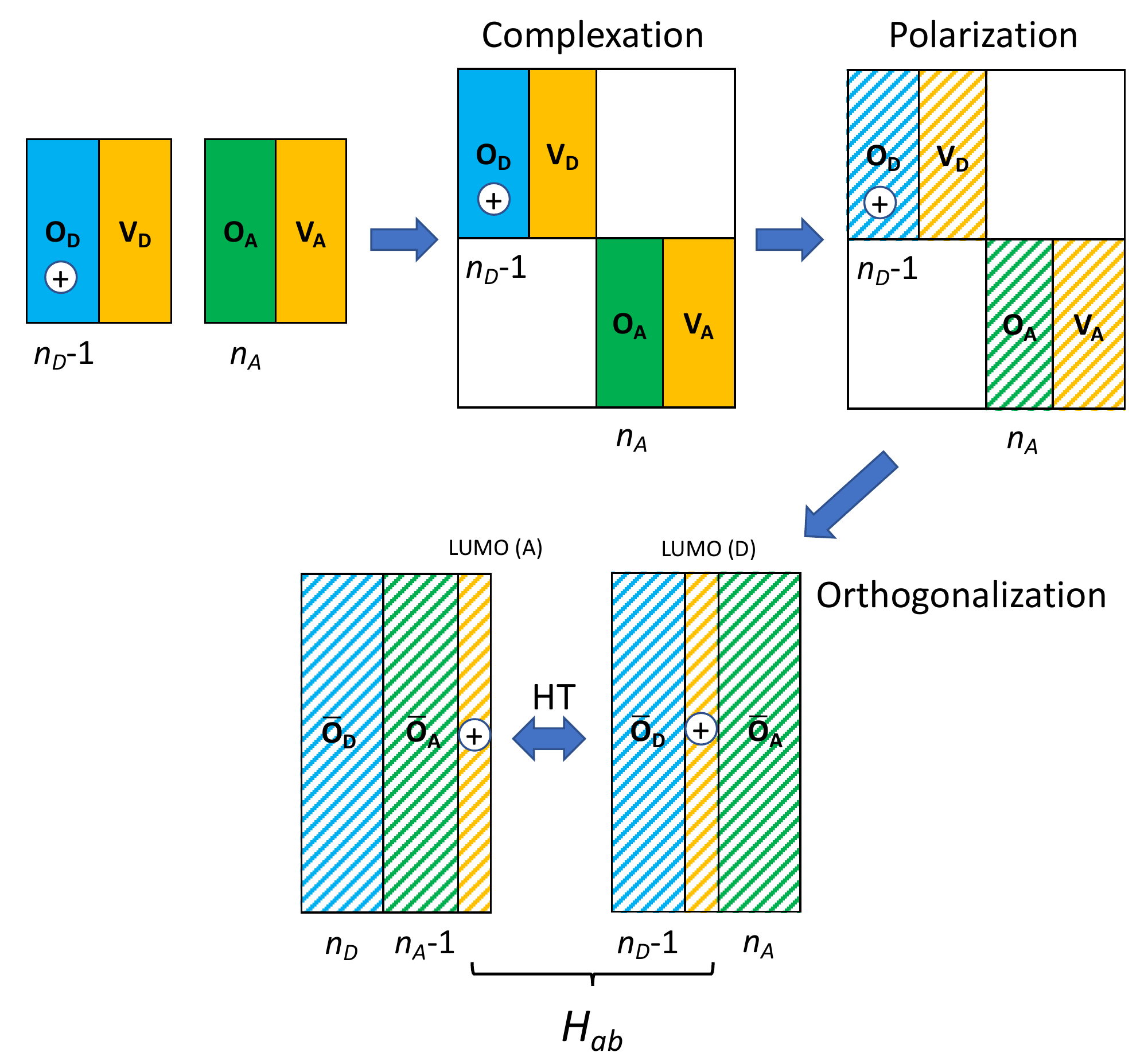}
    \caption{The ALMO(FODFT) procedure, including: (i) SCF calculations for fragments $D^+$ and $A$; (ii) construction of the reactant state $\ket{D^+A}$ from fragment orbitals; (iii) variational relaxation of the ALMO state; (iv) orthogonalization of the occupied orbitals and evaluation of $H_{ab}$ using the FODFT scheme.  }
    \label{fig:almo_fodft}
\end{figure}

The entire procedure of the ALMO(FODFT) approach is illustrated in Fig.~\ref{fig:almo_fodft}. From the construction of $\ket{\bar{\psi}_a}$ and $\ket{\bar{\psi}_b}$, one can see that the orbitals are only variationally optimized within one of the diabats (the reactant state $\ket{\bar{\psi}_a}$ in the above example) and are thus not optimal for the other. In addition, the KS Fock operator, used to evaluate $H_{ab}$, is constructed exclusively from the occupied orbitals in the chosen diabat. As a consequence, this scheme will yield the unphysical result $H_{ab} \ne H_{ba}$ for asymmetric systems. This is a problem inherited from the FODFT method introduced in Ref.~\citenum{Schober2016}. One could attempt to restore the symmetry by an approach such as averaging $H_{ab}$ and $H_{ba}$ using the two different choices of the optimized diabat. However, one can avoid this symmetry-breaking problem by employing the ALMO(MSDFT) approach as discussed below.

\subsection{ALMO(MSDFT) approaches} \label{subsec:almo_msdft}

The ALMO(MSDFT) approach avoids the symmetry problems of ALMO(FODFT) by variationally optimizing both the donor and acceptor ALMO states and then using MSDFT\cite{Cembran2009,Ren2016} to evaluate the electronic coupling between them. 
To achieve this, given two diabatic states $\ket{\psi_a}$ and $\ket{\psi_b}$, one can construct a diabatic Hamiltonian:
\begin{equation}
    \mathbf{H}^{\prime} = 
    \begin{pmatrix}
    H^{\prime}_{aa} & H^{\prime}_{ab} \\
    H^{\prime}_{ba} & H^{\prime}_{bb}
    \end{pmatrix}.
\end{equation}
Note that the off-diagonal element, $H^{\prime}_{ab}$, of the matrix above cannot be directly taken as the diabatic coupling if states $\ket{\psi_a}$ and $\ket{\psi_b}$ are non-orthogonal to each other, which is the case for two ALMO states as defined in Eqs.~(\ref{eq:almo_donor}) and (\ref{eq:almo_acceptor}). For systems comprised of an arbitrary number of diabatic states the couplings can be obtained from the off-diagonal elements of the Hamiltonian transformed into the L\"owdin-orthogonalized basis
\begin{equation}
\mathbf{H} = \mathbf{S}^{-1/2} \mathbf{H}^{\prime} \mathbf{S}^{-1/2},
\end{equation}
where $\mathbf{S}$ is the overlap between diabatic states. For two diabats $\ket{\psi_a}$ and $\ket{\psi_b}$, the overlap matrix element is given by
\begin{equation}
    S_{ab} = \Braket{\psi_a | \psi_b} = \mathrm{det}[(\mathbf{C}^{(a)}_{\text{o}})^{T} \mathbf{S} \mathbf{C}^{(b)}_{\text{o}}].
    \label{eq:state_overlap}
\end{equation}
In the two-state case considered above, the L\"owdin orthogonalization procedure yields the following expression for the diabatic coupling
\begin{equation}
    H_{ab} = \frac{1}{1-S_{ab}^2} \left\vert H^{\prime}_{ab} - \frac{H^{\prime}_{aa} + H^{\prime}_{bb}}{2} S_{ab} \right\vert,
    \label{eq:noci_coupling}
\end{equation}

The original MSDFT approach\cite{Cembran2009} provided a protocol to construct the diabatic Hamiltonian $\mathbf{H}^{\prime}$. In this protocol, the diagonal elements are given by the KS energies of the two diabats defined in Eq.~(\ref{eq:ks_energy}):
\begin{equation}
    H^{\prime}_{aa} = E_{a}^{\text{KS}}[\mathbf{P}^{(a)}], \quad H^{\prime}_{bb} = E_{b}^{\text{KS}}[\mathbf{P}^{(b)}]
\end{equation}
where $\mathbf{P}^{(a)}$ and $\mathbf{P}^{(b)}$ are 1PDMs associated with diabats $\ket{\psi_a}$ and $\ket{\psi_b}$, respectively. The approximation for the off-diagonal elements is theoretically more challenging and one approach suggested in Ref.~\citenum{Cembran2009} is
\begin{equation}
    H^{\prime}_{ab} = S_{ab} \left[V_{\text{nn}} + \mathbf{P}_{ab}\cdot\mathbf{h} + \frac{1}{2} \mathbf{P}_{ab} \cdot \mathrm{\RNum{2}} \cdot \mathbf{P}_{ab} + \frac{1}{2}(\Delta E_{a}^{\text{c}} + \Delta E_{b}^{\text{c}})  \right],
    \label{eq:msdft1}
\end{equation}
where the transition density matrix $\mathbf{P}_{ab}$ is
\begin{equation}
    \mathbf{P}_{ab} = \mathbf{C}_{\text{o}}^{(a)} \left[(\mathbf{C}_{\text{o}}^{(b)})^{T} \mathbf{S} \mathbf{C}_{\text{o}}^{(a)} \right]^{-1} (\mathbf{C}_{\text{o}}^{(b)})^{T}.
    \label{eq:trans_den}
\end{equation}
The first three terms in the square brackets in Eq.~(\ref{eq:msdft1}) can be derived by treating the two KS determinants as those obtained from Hartree-Fock (HF) calculations. The last term is thus intended to account for the contribution from XC, and is defined as the average of the difference between the KS and HF energies calculated from the same 1PDM for each diabat:
\begin{subequations}
\begin{align}
    \Delta E_{a}^{\text{c}} &= E^{\text{KS}}_{a} [\mathbf{P}^{(a)}] - E^{\text{HF}}_{a} [\mathbf{P}^{(a)}], \\
    \Delta E_{b}^{\text{c}} &= E^{\text{KS}}_{b} [\mathbf{P}^{(b)}] - E^{\text{HF}}_{b} [\mathbf{P}^{(b)}].
\end{align}
\end{subequations}

The approximation given by Eq.~(\ref{eq:msdft1}) provides a practical approach to calculate the off-diagonal element $H^{\prime}_{ab}$ and was therefore later suggested for the evaluation of diabatic couplings between two ALMO states.\cite{Ren2016, Guo2018} However, using Eq.~(\ref{eq:msdft1}) as the expression for $H^{\prime}_{ab}$ and substituting it into Eq.~(\ref{eq:noci_coupling}) leads to the cancellation of all XC contributions to the diagonal and off-diagonal elements of $\mathbf{H}^{\prime}$ in the evaluation of $H_{ab}$. In other words, the approximation adopted in the original MSDFT approach reduces to treating the two diabats calculated from KS-DFT as determinants constructed by HF orbitals and then using the exact $\hat{H}$ to evaluate the matrix elements. This neglect of the XC contribution to the diabatic coupling, is likely the origin of the systematic overestimation of $\vert H_{ab} \vert$ demonstrated in Ref.~\citenum{Guo2018} and Sec.~\ref{sec:results}.

To incorporate the XC contribution to the diabatic coupling, we employ the following alternative approach: instead of using the difference between KS and HF energies for each diabat, we use the XC energy evaluated from the symmetrized transition density matrix, $\tilde{\mathbf{P}}_{ab}$, to account for the XC contribution to the off-diagonal element:
\begin{equation}
     H^{\prime}_{ab} = S_{ab} \left[V_{\text{nn}} + \mathbf{P}_{ab}\cdot\mathbf{h} + \frac{1}{2} \mathbf{P}_{ab} \cdot \mathrm{\RNum{2}} \cdot \mathbf{P}_{ab} + E_{\text{xc}}[\tilde{\mathbf{P}}_{ab}]  \right],
    \label{eq:msdft2}
\end{equation}
where the symmetrized transition density matrix is
\begin{equation}
    \tilde{\mathbf{P}}_{ab} = \frac{1}{2} (\mathbf{P}_{ab} + \mathbf{P}_{ba}).
    \label{eq:symm_den}
\end{equation}
An advantage of Eq.~(\ref{eq:msdft2}) is that when $a = b$, one recovers the ALMO diabatic state energies given by Eq.~(\ref{eq:ks_energy}).

Although use of the transition density was suggested as a potential possibility in the original MSDFT paper\cite{Cembran2009} to calculate the diabatic Hamiltonian formed by the reactant and product states in nucleophilic substitution reactions, it does not appear to have been previously pursued for calculating diabatic couplings\cite{Ren2016,Guo2018} despite the advantages listed above. We thus systematically demonstrate the advantages of such an approach. In particular, by employing the symmetrized transition density as input to the XC functional in Eq.~(\ref{eq:symm_den}), we circumvent the potential symmetry-breaking problem in the evaluation of the off-diagonal elements of $\mathbf{H}^{\prime}$. To distinguish the approach given by Eqs.~(\ref{eq:msdft2}) and (\ref{eq:symm_den}) from the original MSDFT scheme that uses Eq.~(\ref{eq:msdft1}), we refer to the former as ALMO(MSDFT2).

\section{Implementation and Computational Details} \label{sec:comput_detail}

We implemented POD, FODFT, and all ALMO-based diabatization schemes in a development version of the Q-Chem 5.2 package.\cite{Shao2015}  For POD, consistent with Ref.~\citenum{Yang2018} we start with a full SCF calculation for the closed-shell reference system to prepare the canonical KS orbitals for both ET and HT cases. Although implementation of POD using SCF solutions for charged reference systems has also been reported\cite{Futera2017}, we find that this approach yields too large $\vert H_{ab}\vert$ values when employing hybrid functionals. With POD we report $H_{ab}$ as the Fock matrix element between the HOMOs/LUMOs of the donor and acceptor blocks (both in the neutral state) for HT/ET.

We investigate three previously suggested methods of constructing FODFT fragments orbitals and the Fock operators in this work: (i) FODFT($2n$)@$DA$,\cite{Senthilkumar2003} (ii) FODFT($2n\mathrm{-}1$)@$DA$~/~FODFT($2n\mathrm{+}1$)@$D^-A^-$,\cite{Oberhofer2012} and (iii) FODFT($2n\mathrm{-}1$)@$D^+A$~/~FODFT($2n\mathrm{+}1$)@$D^-A$\cite{Schober2016}. Here the fragment symbols ($D$ and $A$) following ``@" specify whether neutral or charged fragments are employed to prepare the fragment orbitals and the number in the parentheses indicates the number of occupied fragment orbitals that are used to construct the KS Fock operator ($\hat{f}_{\text{KS}}$). For approaches (ii) and (iii), the fragment orbitals are prepared with different reference states for HT and ET, and the corresponding methods are denoted as FODFT($2n\mathrm{-}1$) and FODFT($2n\mathrm{+}1$), respectively. Note that for approaches (i) and (iii), the total number of electrons in the fragment SCF calculations are consistent with the number of occupied orbitals that contribute to $\hat{f}_{\text{KS}}$, while for approach (ii) one of the fragment occupied orbitals needs to be excluded when constructing the Fock operator. We refer the reader to Ref.\citenum{Schober2016} for a detailed discussion about the different flavors of FODFT approaches. Unless otherwise specified, for ET we report the coupling between the $(n_D/2+1)$th $\alpha$ MO of the donor and the $(n_A/2+1)$th $\alpha$ MO of the acceptor, and for HT we calculate the coupling between the $(n_D/2)$th and the $(n_A/2)$th $\beta$ MOs of the donor and acceptor fragments, respectively.

The ALMO-based diabatic states are obtained using the unrestricted ``SCF for molecular interaction" (SCF-MI) procedure,\cite{Horn2013} which was originally developed for energy decomposition analysis. The ALMO(FODFT) approach resembles FODFT($2n\mathrm{-}1$)@$D^+A$~/~FODFT($2n\mathrm{+}1$)@$D^-A$ since the charged fragment is explicitly considered in the construction of diabatic wavefunctions and they both assume that ET/HT occurs between a specific pair of fragment orbitals (see Fig.~\ref{fig:almo_fodft}).

In contrast to ALMO(FODFT), the MSDFT based approaches couple KS determinants directly and thus do not require any specification of donor/acceptor orbitals. For MSDFT and MSDFT2 the off-diagonal elements of the diabatic Hamiltonian were computed using Eq.~(\ref{eq:msdft1}) and (\ref{eq:msdft2}), respectively. In both cases the Q-Chem routines for non-orthogonal configuration interaction (NOCI)\cite{Thom2009,Sundstrom2014} were used to evaluate the non-XC terms. For MSDFT2, the additional XC contribution (the last term in Eq.~\ref{eq:msdft2}) was then incorporated by evaluating the XC energy using the symmetrized transition density matrix (Eq.~\ref{eq:symm_den}).
In our implementation of MSDFT we first orthogonalize the occupied orbitals of each ALMO state and then used the generalized Slater-Condon rules\cite{Amos1961,Thom2009} to evaluate the non-XC terms. Since the orthogonalization of occupied orbitals of each spin separately does not modify the energy and 1PDM associated with each ALMO state, our implementation of these non-XC terms is equivalent to those previously employed based on the 1PDM and transition density matrix using Eqs.~\ref{eq:almo_den} and \ref{eq:trans_den}, respectively. 

Our CDFT calculations utilized the implementation of CDFT-CI\cite{Wu2007} in Q-Chem 5.2, in which the off-diagonal element of the diabatic Hamiltonian is computed as
\begin{subequations}
\begin{align}
    H^{\prime}_{ab} &= (E_b + V_{b} N_{b}) \braket{\psi_a | \psi_b} - V_{b} \braket{\psi_a | \hat{w} | \psi_b} \\
    H^{\prime}_{ba} &= (E_a + V_{a} N_{a}) \braket{\psi_b | \psi_a} - V_{a} \braket{\psi_b | \hat{w} | \psi_a},
\end{align} \label{eq:cdft_offdiag}
\end{subequations}
where $E_a$, $E_b$ are the KS energies of two CDFT diabats, $V_a$, $V_b$ are the Lagrangian multipliers at the convergence of the constrained SCF calculations, and $N_a$, $N_b$ are the constrained values in CDFT calculations. In the case of ET/HT, we choose $N_a$ and $N_b$ to be the charge populations on the donor fragment for the reactant and product states, respectively. $\braket{\psi_a | \psi_b}$ denotes the overlap between two diabats and $\hat{w}$ is a one-body partition operator (also known as weighting function) that is determined by the employed population scheme. As in the original CDFT-CI scheme,\cite{Wu2007} we employ the Becke partition approach\cite{Becke1988} and determine the constrained value for each diabat by projecting the ``promolecule" density (the sum of non-interacting fragment densities) onto the weighting function for the preparation of CDFT states in this work. Owing to the approximations made in the derivation of Eq.~(\ref{eq:cdft_offdiag}), in general $H^{\prime}_{ab} \neq H^{\prime}_{ba}$.\cite{Wu2006} To avoid this symmetry issue we follow the procedure used in previous CDFT studies\cite{Wu2006b, Wu2006, Wu2007, VanVoorhis2010} and use the average of $H^{\prime}_{ab}$ and $H^{\prime}_{ba}$ as the off-diagonal element of the diabatic Hamiltonian. We then obtain the diabatic coupling ($H_{ab}$) using Eq.~(\ref{eq:noci_coupling}). 

Except where otherwise stated all our DFT-based calculations for diabatic couplings were performed with the 6-31+G(d) basis set\cite{Hehre1972,Frisch1984} on a (99, 590) grid (99 radial shells with 590 Lebedev points in each). To examine how the performance of each scheme is influenced by the choice of basis set, we also generated benchmark results for the HAB11 HT dataset with three other double- or triple-$\zeta$ basis sets: def2-SVPD, def2-TZVPD,\cite{Weigend2005, Rappoport2010} and aug-cc-pVTZ\cite{Dunning1989,Woon1993}. As an alternative approach to assess, we also computed diabatic couplings using time-dependent density functional theory (TDDFT)\cite{Runge1984, Burke2005} as well as its spin-flip variant \cite{Shao2003} (based on Eq.~\ref{eq:half_gap} below), which utilize unrestricted doublet and quartet SCF solutions as references, respectively. Note that one needs to avoid symmetry breaking in obtaining these unrestricted SCF solutions. Finally, we performed EOM-IP-CCSD\cite{Stanton1994} calculations to generate or appraise the reference $\vert H_{ab} \vert$ values for some of the HT complexes, which start from neutral, closed-shell reference states.

To measure the performance of different schemes on benchmark datasets, we use the mean unsigned relative error (MURE) that is calculated as $\text{MURE} = (\sum_{i=1}^{n} \vert y_i - y_{i, \text{ref}} \vert / y_{\text{ref}})/n$ and the and root-mean-square relative error (RMSRE) which is $\text{RMSRE} = [\sum_{i=1}^{n} (\vert y_i - y_{i, \text{ref}} \vert / y_{\text{ref}})^2/n]^{1/2}$, where $n$ is the total number of points in the dataset. In the Supporting Information (SI), we also show mean signed relative errors [$\text{MSRE} = (\sum_{i=1}^{n} (y_i - y_{i, \text{ref}}) / y_{\text{ref}})/n$] of different approaches, whose comparison against MURE can be used to detect systematic underestimation or overestimation of $\vert H_{ab} \vert$ values.

\section{Results} \label{sec:results}

\begin{figure*}[t!]
    \centering
    \includegraphics[width=0.85\textwidth]{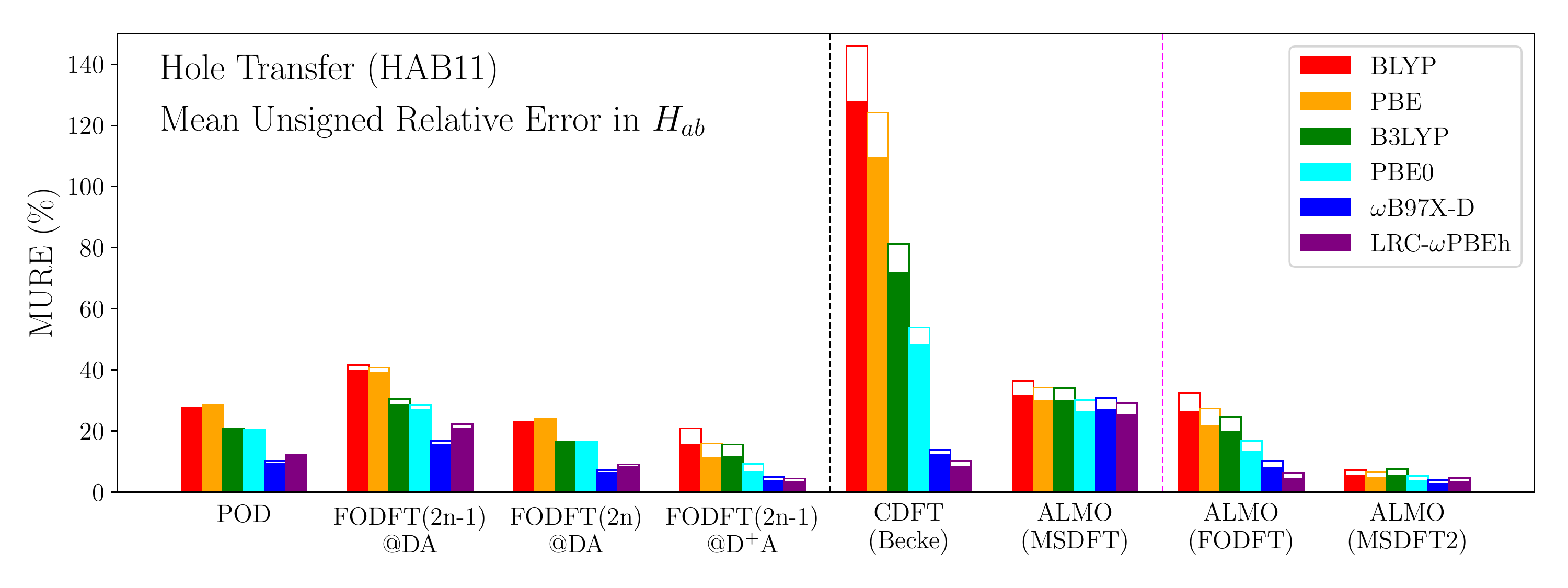}
    \caption{Performance of the diabatization schemes for the diabatic couplings ($\vert H_{ab} \vert$) in the HAB11 hole transfer dataset. The unfilled bars indicate the difference between the MURE and the RMSRE. Methods on the right of the black dashed line are ones that can produce variationally optimized diabatic states, and those on the right of the magenta dashed line are ones proposed in this work.}
    \label{fig:hab11}
\end{figure*}

We examine the accuracy of DFT-based diabatization schemes on electronic couplings for HT and ET using two previously introduced benchmark sets, each consisting of symmetric dimers of organic molecules. For HT we benchmark against the HAB11 dataset,\cite{Kubas2014} which consists of reference $\vert H_{ab} \vert$ values for 11 homo-dimers of small-to-medium sized organic molecules (e.g.~ethylene and thiophene). For ET we benchmark against HAB7-,\cite{Kubas2015} which contains references values for 7 medium-to-large organic homo-dimers (e.g.~tetracene and porphin). 

Because of the symmetry of the dimer systems, the two lowest energy diabatic states are degenerate, and the electronic coupling between them can therefore be calculated as
\begin{equation}
    \vert H_{ab} \vert = \frac{1}{2} (E_1 - E_0),
    \label{eq:half_gap}
\end{equation}
where $E_0$ and $E_1$ are the energies of the adiabatic ground and first excited states, respectively, which can be obtained using high-level multireference electronic structure methods. In the HAB11 dataset these have been calculated previously\cite{Kubas2014} using MRCI+Q and NEVPT2 and for HAB7- using SCS-CC2\cite{Kubas2015}. For each system, the $\vert H_{ab} \vert$ values at four distinct distances (3.5, 4.0, 4.5, and 5.0 \AA) are provided, enabling one to also examine the exponential decay of $\vert H_{ab} \vert$ with respect to intermolecular distance 
\begin{equation}
    \vert H_{ab}(d) \vert = A \exp(-\beta d/2),
\end{equation}
which is the expected asymptotic decay of the diabatic coupling. Here $\beta$ is the exponential decay constant that is to be examined, $d$ is the intermolecular distance, and $A$ is a pre-exponential factor. Given the values of $\vert H_{ab} \vert$ at four distances, we obtain $\beta$ by performing a linear regression of $\log \vert H_{ab} \vert$ against $d$.

To assess the performance of each diabatization scheme when paired with different types density functionals, we generated benchmark results at three levels of density functional theory: pure GGA (BLYP\cite{Becke1988b, Lee1988}, PBE\cite{Perdew1996}), global hybrid (B3LYP\cite{Becke1993}, PBE0\cite{Adamo1999}), and range-separated hybrid ($\omega$B97X-D\cite{Chai2008}, LRC-$\omega$PBEh\cite{Rohrdanz2009}). 

\subsection{Hole transfer dataset}

Figure \ref{fig:hab11} shows the errors in $\vert H_{ab} \vert$ of the DFT-based diabatization schemes for the HAB11 HT dataset. As a general trend, the range-separated hybrid (RSH) functionals produce more accurate results than the pure GGAs and global hybrids, while the advantage of RSH over other functionals varies from method to method. Among the diabatization methods that do not provide variationally optimized diabats (those on the left of the black dashed line in the figure), FODFT($2n\mathrm{-}1$)@$D^+A$ yields the most accurate diabatic couplings. This should be expected since it explicitly accounts for the positive charge on the donor fragment when calculating the fragment orbitals. POD and FODFT($2n$)@$DA$, which both employ a neutral, closed-shell reference state, also yield fairly accurate $H_{ab}$ values, especially when using RSH functionals. FODFT($2n\mathrm{-}1$)@$DA$ gives the least accurate results among the three flavors of FODFT approaches, which is consistent with previous results on this dataset showing that this approach systematically underestimates diabatic couplings\cite{Kubas2014}. This is likely due to the inconsistency of the number of electrons in the preparation of fragment orbitals, which correspond to $n_D + n_A$ electrons for a neutral system, and in the construction of the global Fock matrix, where one electron is removed. 

Turning to the methods that can produce variationally optimized diabatic states (on the right of the black dashed line), as previously observed\cite{Kubas2014, Kim2017} the performance of CDFT is highly sensitive to the choice of XC functional with the MURE dropping roughly 15 times upon going from the BLYP GGA functional to the best-performing RSH (LRC-$\omega$PBEh). In contrast to CDFT, the original ALMO(MSDFT) approach uniformly overestimates $\vert H_{ab} \vert$ (see Table S1 in the SI for the signed errors) regardless of the functional employed. ALMO(MSDFT) shows a marginal improvement in the MURE upon going from BLYP (32\%) to LRC-$\omega$PBEh (26\%), which is in agreement with the trend observed in Ref.~\citenum{Guo2018} that the error of MSDFT on HAB11 reduces upon going from PBE to PBE0 and finally to PBEC (100\% HF exchange + PBE correlation). This can be explained by our observation in Sec.~\ref{subsec:almo_msdft} that the original MSDFT approach in fact uses the HF operator to compute both the diagonal and off-diagonal elements of the diabatic Hamiltonian. Therefore, functionals with a higher percentage of HF exchange are likely to yield more internally consistent results.

\begin{figure*}[t!]
\centering
\includegraphics[width=0.9\textwidth]{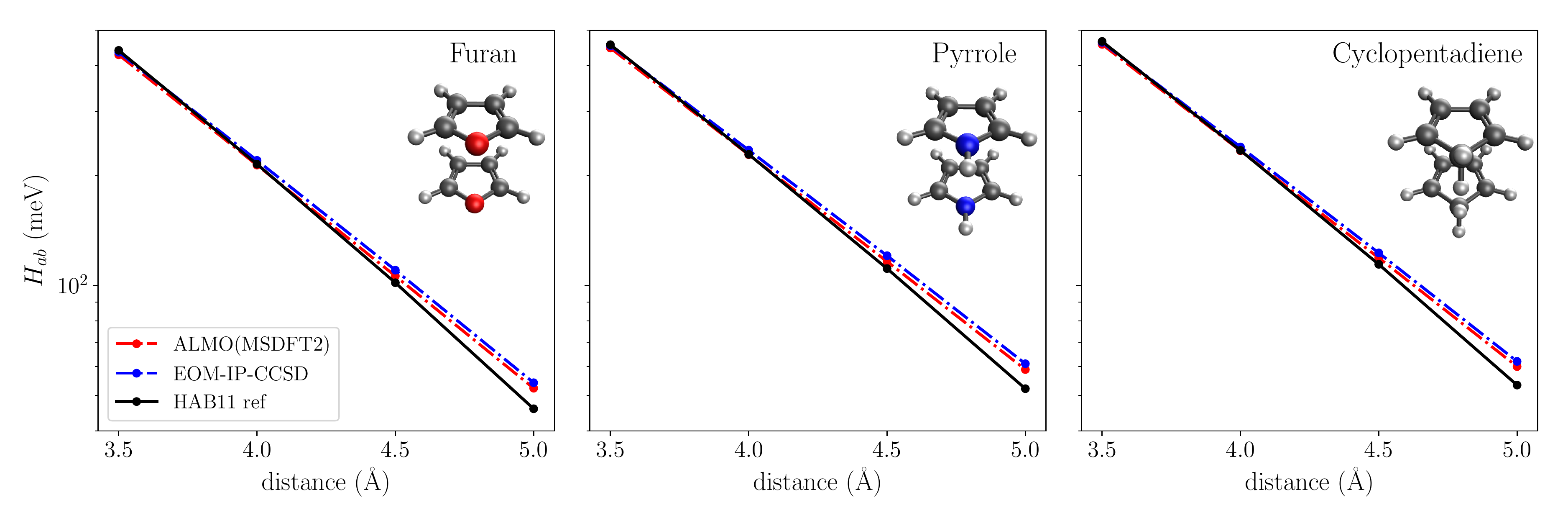}
\caption{Comparison of the reference values in the HAB11 dataset to those obtained from the EOM-IP-CCSD electronic structure method and from ALMO(MSDFT2) (using LRC-$\omega$PBEh) for the distance dependence of the $\vert H_{ab} \vert$ values for furan, pyrrole, and cyclopentadiene. The energy splitting between the first and second IP states (from the closed-shell reference) is utilized to calculate the diabatic coupling (similar to Eq.~\ref{eq:half_gap}) except for the pyrrole dimer at 3.5 \AA, for which the gap between the first and third states is used due to the switching of state ordering.}
\label{fig:hab11_eom}
\end{figure*}

Moving to the methods proposed in this work (those on the right of the magenta line), the ALMO(FODFT) approach shows a similar functional dependence to the related FODFT method (FODFT($2n\mathrm{-}1)$@$D^+A$), while offering the advantage of producing variationally optimized diabatic states. Its MURE reduces significantly when paired with the RSH functionals, dropping from 26.4\% with BLYP to 4.8\% with LRC-$\omega$PBEh. Moreover, the MSREs of this approach are consistently positive for all tested functionals, indicating its systematic overestimation of the $\vert H_{ab} \vert$ values (see Table S1 in the SI). In contrast to this and all the other methods, ALMO(MSDFT2) gives diabatic couplings in very good agreement when combined with \textit{any} level of XC functional tested. Even for lowest level pure GGA functionals, BLYP and PBE, the MUREs for ALMO(MSDFT2) are only 5.8\% and 5.1\%, respectively, which are $\sim$6 times lower than those arising from ALMO(MSDFT) with a GGA functional and $\sim$20 times lower than those from CDFT. Indeed, even when ALMO(MSDFT) and CDFT are used with the best-performing RSH functional tested (LRC-$\omega$PBEh), their MUREs of 25.6\% and 8.5\% still do not surpass the results obtained using ALMO(MSDFT2) with just a GGA functional. Only the FODFT($2n\mathrm{-}1)$@$D^+A$ and ALMO(FODFT) methods with RSHs give marginally ($\sim 2$\%) lower MUREs than the ALMO(MSDFT2) GGA results. Combining ALMO(MSDFT2) with the RSH $\omega$B97X-D functional reduces the MURE to 3.1\%, which is better than from any other method tested.

\begin{figure}[h!]
\centering
\includegraphics[width=0.48\textwidth]{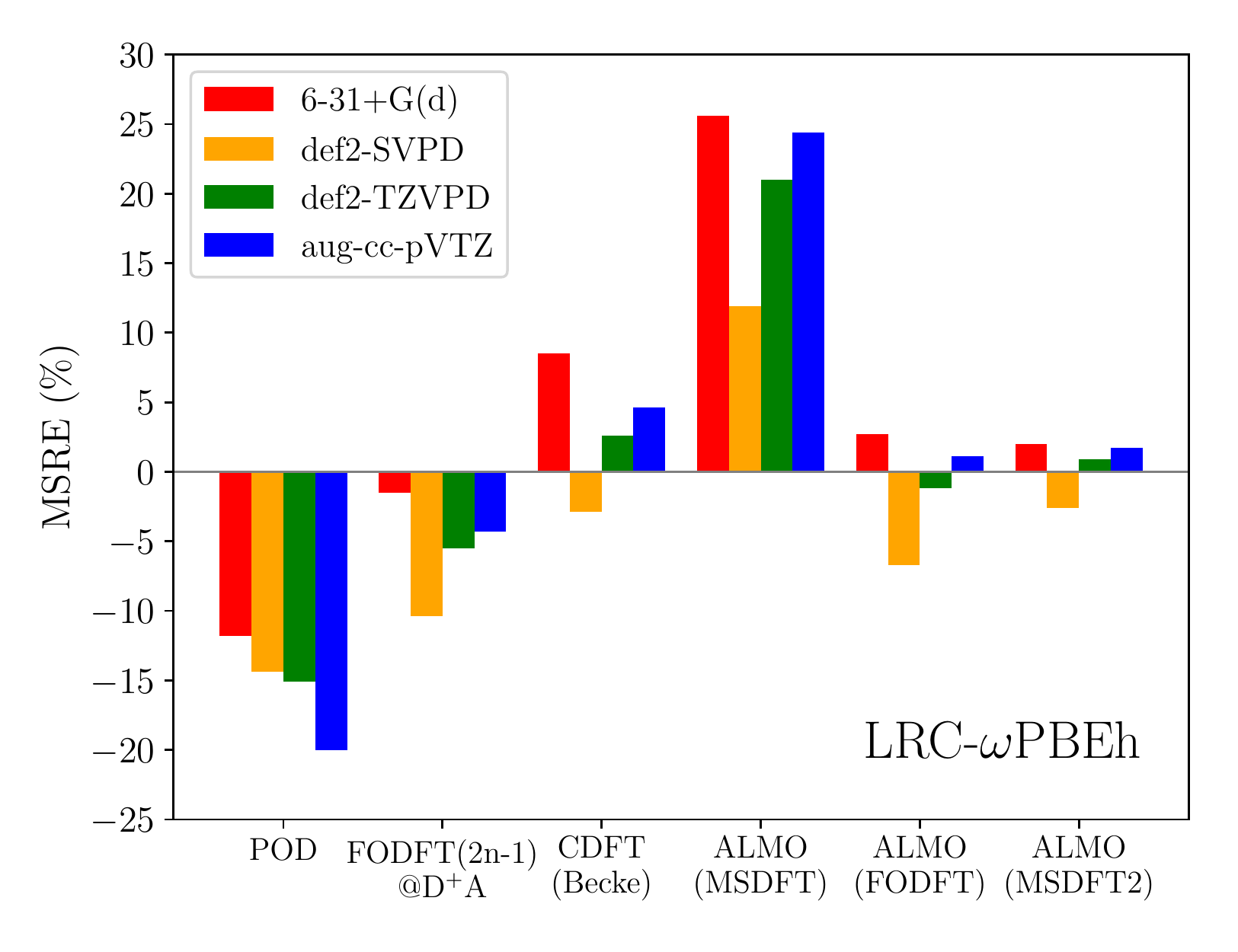}
\caption{Effect of basis set on the MSREs of the diabatization schemes in $\vert H_{ab} \vert$ values for the hole transfer (HAB11) dataset. The calculations are performed with the LRC-$\omega$PBEh functional and four different basis sets: 6-31+G(d), def2-SVPD, def2-TZVPD, and aug-cc-pVTZ.}
\label{fig:basis_trans_hab11}
\end{figure}

Given the excellent performance of ALMO(MSDFT2), it is worth examining the results for specific molecules in the HAB11 dataset. Doing this (see the original data available in the SI), one can observe that while the relative error in ALMO(MSDFT2) for most molecules is under 5\%, by far the largest errors, which range from 12\% to 14\%, arise for the furan, pyrrole, and cyclopentadiene dimers at the largest intermolecular separation in the HAB11 dataset (5 \AA). However, as shown in Fig.~\ref{fig:hab11_eom}, the previously published reference values for the diabatic coupling at this distance deviate from the expected asymptotic behavior. Hence, to provide alternative high-level reference values for these systems, we performed EOM-IP-CCSD calculations. While EOM-IP-CCSD gives almost identical diabatic couplings at the two shorter distances to the MRCI+Q reference in the HAB11 dataset, at the two longer distances it gives larger values of the coupling, which agree much better with the physically expected exponential decay. The EOM-IP-CCSD results are also much closer to the ALMO(MSDFT2) results with a difference of only 4\%, which is more consistent with the errors observed for other systems. This indicates that the couplings reported for these systems at the longer distances in the HAB11 reference data might be underestimated.

Due to the potential underestimation of $\vert H_{ab} \vert$ at long distances, the reference value for the exponential decay constants, $\beta$, for the HT complexes in HAB11 might also be affected. Figure S1 in the SI shows the MURE in the resulting exponential decay constants, $\beta$, for the HT complexes in HAB11. The performance and trends in the ability of the DFT-based diabatization schemes to obtain $\beta$ are in most cases similar to that observed for $\vert H_{ab} \vert$. Two methods that perform noticeably better in their prediction of $\beta$ than $\vert H_{ab} \vert$ are POD and FODFT($2n$)@$DA$. This indicates that these methods are in error by a roughly constant factor at all distances, which is also consistent with their extremely small gaps between MURE and RMSRE in Fig.~\ref{fig:hab11}. ALMO(MSDFT2) again gives small errors across the entire range of functionals tested, with MUREs in $\beta$ ranging from 2.2\% ($\omega$B97X-D) to 4.3\% (BLYP).

\begin{figure*}[t!]
    \centering
    \includegraphics[width=0.85\textwidth]{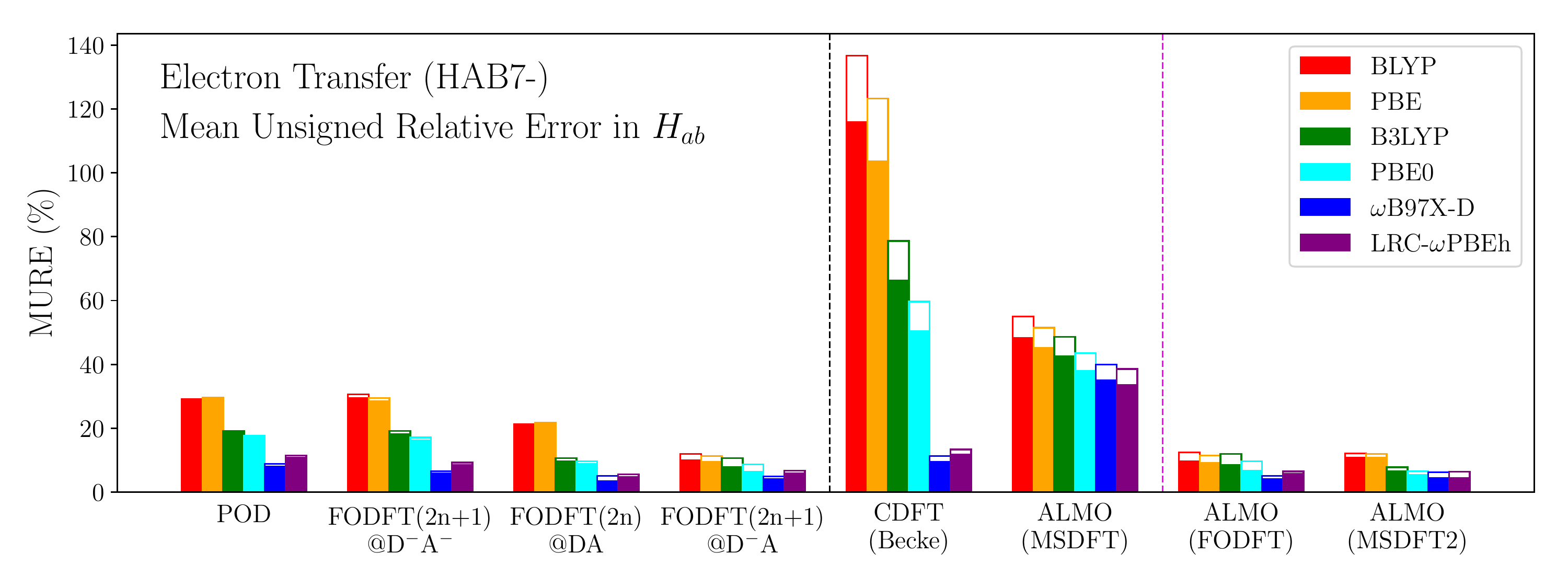}
    \caption{Performance of the diabatization schemes for the diabatic couplings ($\vert H_{ab} \vert$) in the HAB7- electron transfer dataset.
    The other plotting details are the same as in Fig.~\ref{fig:hab11}.}
    \label{fig:hab7}
\end{figure*}

To assess the robustness of the DFT-based diabatization schemes with respect to basis set size, we show their performance based on the MSRE when paired with the LRC-$\omega$PBEh functional using four different basis sets in Fig.~\ref{fig:basis_trans_hab11}. The MSRE reports on the systematic over- or underestimation of a given method caused by basis set changes. In particular, it is well-known that the charge-transfer energy given by the original ALMO-based energy decomposition analysis,\cite{Khaliullin2007} which is defined as the energy difference between the lower-energy diabatic state and the fully delocalized ground adiabatic state, can be sensitive to the basis size\cite{Lao2016, Mao2018a}. However, the diabatic couplings obtained using ALMO(MSDFT2) shown in Fig.~\ref{fig:basis_trans_hab11} do not suffer from this problem, with MSRE confined to a tight range between -2.6\% to 2.0\% and the MURE between 2.9\% to 3.5\% across the investigated basis sets. This demonstrates the transferability of the accuracy of ALMO(MSDFT2) across basis sets. In contrast, the results obtained from other methods show a more substantial dependence on the basis set. In particular, the POD method underestimates the diabatic couplings even more when the basis set becomes larger and more diffuse, while the accuracy of CDFT, in contrast, improves when one utilizes a triple-$\zeta$ basis set such as def2-TZVPD. ALMO(FODFT) also shows good accuracy when paired with triple-$\zeta$ basis sets but the errors span a larger range compared to those of ALMO(MSDFT2). Finally, as shown in SI Fig.~S3, one should note that at least one set of diffuse functions in the basis set is required to reproduce the exponential decay behavior at the longest distances in the HAB11 dataset, and that using a smaller basis set, such as 6-31G(d), can cause significant underestimation of $\vert H_{ab} \vert$ at long range.

\subsection{Electron transfer dataset}

Figure~\ref{fig:hab7} shows the errors in $\vert H_{ab} \vert$ of the DFT-based diabatization schemes for the HAB7- ET dataset. In general, the performance of these schemes for ET shown in Fig.~\ref{fig:hab7} resemble those for HT in Fig.~\ref{fig:hab11}. With RSH functionals, the POD and all the FODFT approaches yield fairly accurate $H_{ab}$ values with FODFT($2n\mathrm{+}1$)@$D^{-}A$ performing particularly well with all levels of functionals. For the methods that provide variationally optimized diabats, the strong functional dependence of CDFT results and the uniformly overestimated ALMO(MSDFT) diabatic couplings still hold for the ET dataset with the latter giving MUREs of over 30\% even when paired with RSH functionals. The performance of ALMO(FODFT) with different functionals mirrors that of the related FODFT($2n\mathrm{+}1$)@$D^{-}A$ approach. This is in contrast to the HT case, where the FODFT approach without orbital relaxation gives noticeably better results than ALMO(FODFT). ALMO(MSDFT2) produces accurate diabatic couplings with all levels of functionals. The smallest MUREs (4.7\%) are achieved with the $\omega$B97X-D and LRC-$\omega$PBEh RSH functionals, which are very close to the best MUREs achieved by FODFT-based approaches with $\omega$B97X-D of 3.7\% and 4.3\%. With the GGA functionals, the MURE of ALMO(MSDFT2) increases to around 11\%, which is about 1\% larger than the corresponding MUREs of FODFT($2n\mathrm{+}1$)@$D^-A$ and ALMO(FODFT). Similar trends are also observed in the exponential decay rates, $\beta$, as shown in SI Fig.~S2.

\begin{figure}[t!]
    \centering
    \includegraphics[width=0.4\textwidth]{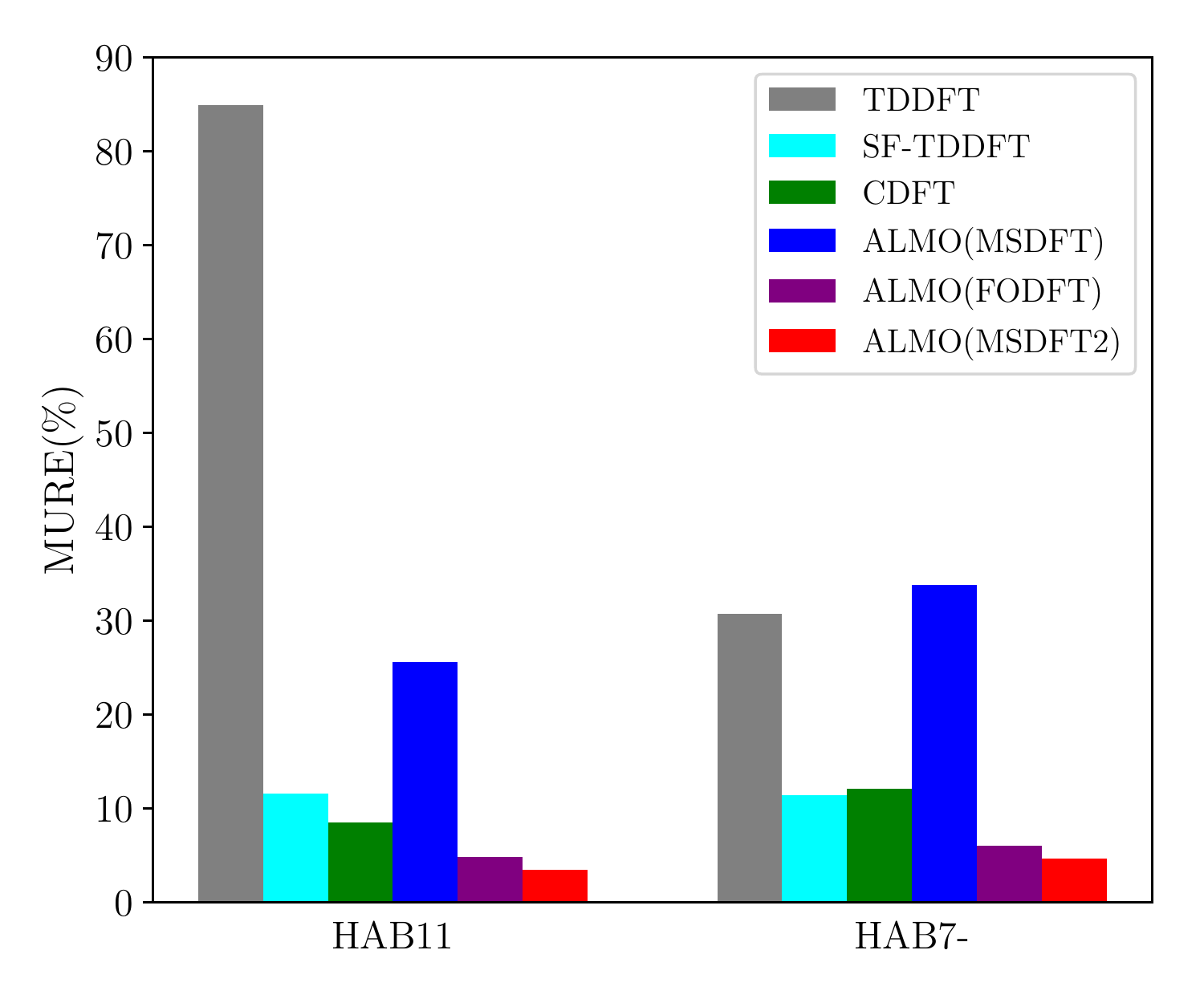}
    \caption{Comparison of the MUREs of TDDFT and SF-TDDFT against those of DFT-based diabatization schemes that construct variationally optimized diabatic states directly [CDFT, ALMO(MSDFT), ALMO(FODFT), and ALMO(MSDFT2)]. All calculations are performed with the LRC-$\omega$PBEh functional and the 6-31+G(d) basis. The results calculated with $\omega$B97X-D are provided in Fig.~S4 in the SI.}
    \label{fig:tddft}
\end{figure}

\begin{figure*}[t!]
    \centering
    \includegraphics[width=\textwidth]{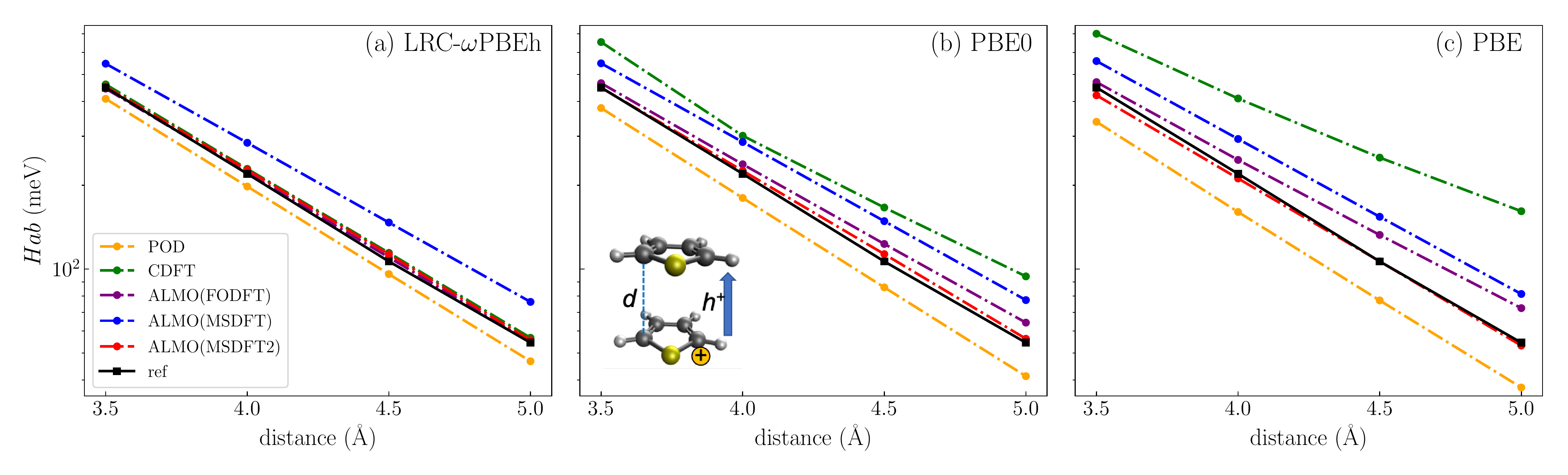}
    \caption{Effect of the tier of exchange-correlation functional used on the performance of the diabatization schemes in capturing the distance dependence of $\vert H_{ab} \vert$ for HT in the thiophene dimer. The calculations are performed with (a) LRC-$\omega$PBEh, (b) PBE0, and (c) PBE functionals. The reference values were obtained from Ref.~\citenum{Kubas2014}. The $y$-axis ($\vert H_{ab} \vert$) is in a logarithmic scale and is shared by all three panels. Analogous results for HT and ET in a perfectly stacked pentacene dimer are shown in SI Fig.~S5.}
    \label{fig:thiophene_dist}
\end{figure*}

\begin{figure*}[t!]
    \centering
    \includegraphics[width=\textwidth]{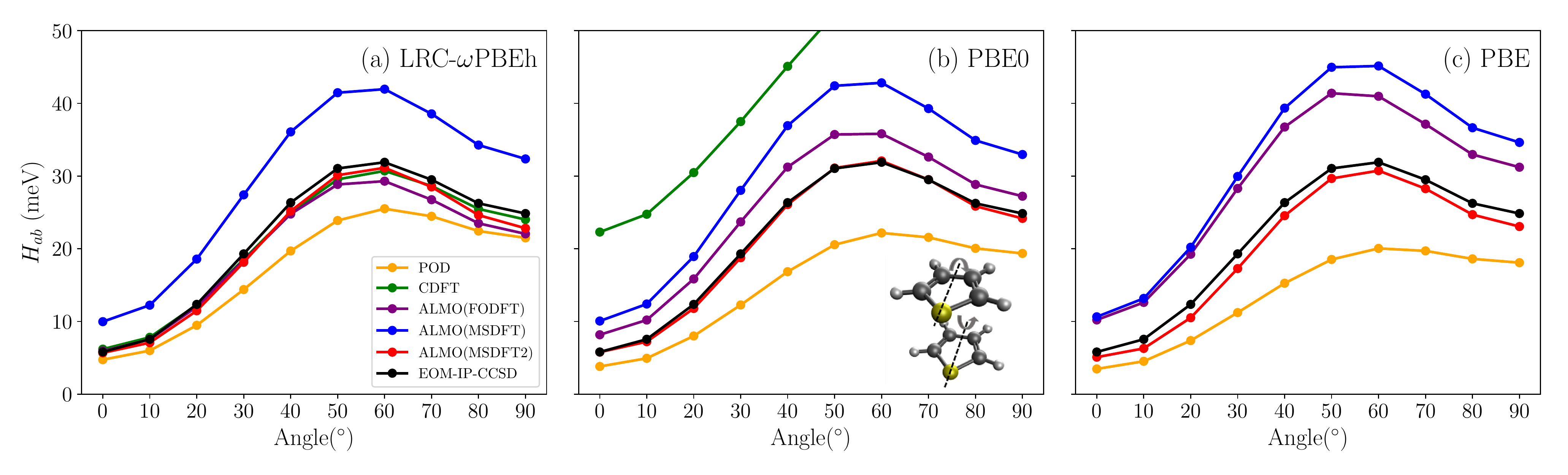}
    \caption{Effect of the tier of exchange-correlation functional used on the performance of the diabatization schemes in capturing the angular dependence of $\vert H_{ab} \vert$ for HT in the thiophene dimer. The calculations are performed with (a) LRC-$\omega$PBEh, (b) PBE0, and (c) PBE functionals using the 6-31+G(d) basis set. The reference values were generated with EOM-IP-CCSD using the same basis. The $y$-axis ($\vert H_{ab} \vert$) is shared by all three panels.}
    \label{fig:thiophene_angle}
\end{figure*}

Another approach to obtain the diabatic coupling from DFT is to use the first excitation energies provided by TDDFT calculations combined with Eq.~(\ref{eq:half_gap}). However, owing to the multireference nature of these symmetric HT and ET complexes, which arises from their degenerate diabatic states, one should not expect conventional TDDFT to yield accurate results. Figure~\ref{fig:tddft} shows that, even with the LRC-$\omega$PBEh RSH functional, the MURE in the diabatic couplings given by TDDFT is 85\% for HT, and over 30\% for ET, with the former being consistent with a previous study on the same dataset where TDDFT calculations were performed with the cc-pVTZ basis\cite{Manna2018}. Although one could use an $\omega$-tuning based on the LRC-$\omega$PBEh or LC-BLYP functional to improve the performance of TDDFT, as suggested in recent studies,\cite{Manna2018,Kitoh2019} this requires modifying the functional for each system studied. Alternatively, one can use spin-flip (SF)-TDDFT\cite{Shao2003} to provide a more balanced description of the ground and first excited adiabatic states.\cite{You2004} As we show in Fig~\ref{fig:tddft}, performing SF-TDDFT on the HAB11 and HAB7- datasets yields MUREs of 11.6\% and 11.4\%, respectively. These errors are a considerable improvement over TDDFT but still significantly exceed the values of 3.5\% and 4.7\% obtained from ALMO(MSDFT2). In addition, while in ALMO(MSDFT2) one only needs to specify the charge and spin of each fragment and hence can be used in a nearly black-box manner, extra caution is required when running SF-TDDFT and TDDFT calculations. This is because symmetry-conserving reference states are required for these latter methods to give physical results, which is challenging in practice since the stable solutions of unrestricted SCF usually break symmetry. Further, unlike CDFT and the ALMO-based approaches, the SF-TDDFT diabats are not variationally optimized with respect to orbital rotations, rendering the calculation of forces nontrivial.

\subsection{Hole transfer in thiophene dimer}

Oligomers and polymers of thiophene play an important role in organic electronics.\cite{Liang2010, Berger2018} Here we examine the performance of different DFT-based schemes in predicting the diabatic couplings for HT in a thiophene dimer. As a representative example, we focus on the face-to-face stacked thiophene dimer shown in Fig.~\ref{fig:thiophene_dist} to illustrate that, while all the diabatization methods except POD and ALMO(MSDFT) give satisfactory couplings and decay rates for this type of system when combined with higher-tier RSH functionals [LRC-$\omega$PBEh in Fig.~\ref{fig:thiophene_dist}(a)], the performance of all except ALMO(MSDFT2) degrades markedly when using lower-tier functionals (PBE0 and PBE in Figs.~\ref{fig:thiophene_dist}(b) and (c), respectively). It is interesting to note that the errors arising from many of these methods is systematic, with the POD approach consistently underestimating the diabatic coupling across the functional hierarchy and worsening with lower-tier functionals. In contrast, all other methods except ALMO(MSDFT2) systematically overestimate the couplings when using the lower-tier global hybrid and pure GGA functionals, with the most notable degradation occurring in the case of CDFT. Of all the methods tested, only ALMO(MSDFT2) shows robust performance across the functional hierarchy, capturing the reference results even when paired with the pure GGA functional (PBE).

To ensure that the results in Fig.~\ref{fig:thiophene_dist} are not specific to the particular geometry chosen, we also examined the change in the diabatic coupling values with respect to the simultaneous rotation of each thiophene molecule away from the perfectly face-to-face stacked configuration (geometries obtained from Ref.~\citenum{Kubas2014}). The rotation is illustrated in the middle panel of Fig.~\ref{fig:thiophene_angle}: at $0^\circ$, the centers of the two molecules are separated by 6.57 \AA; they are then rotated around each molecule's $C_2$ axis with opposite clockwise directions. Figure~\ref{fig:thiophene_angle} shows the results of the same set of diabatization methods with the three PBE-based functionals, as well as the reference values that we calculated using EOM-IP-CCSD. Note that the $y$-axis shared by this set of plots is in a normal, non-logarithmic scale. While most methods are able to capture the change in $\vert H_{ab} \vert$ upon the rotation qualitatively, only ALMO(MSDFT2) shows quantitative agreement with the reference values when the lower-tier functionals PBE0 and PBE [panels (b) and (c) of Fig.~\ref{fig:thiophene_angle}, respectively] are employed. The other approaches yield significantly overestimated or underestimated results when paired with lower-tier functionals, which are consistent with the trends revealed in Fig.~\ref{fig:thiophene_dist}. For instance, despite its excellent agreement with EOM-IP-CCSD when combined with a RSH functional, CDFT significantly overestimates $\vert H_{ab} \vert$ when PBE0 or PBE is used. The performance of these diabatization methods for HT in thiophene is mirrored in the results for both ET and HT in a pentacene dimer, which has been of significant recent interest regarding singlet fission processes, as shown in Fig.~S5 in the SI.

\subsection{Hole transfer between DNA bases}

\begin{figure}[t!]
    \centering
    \includegraphics[width=0.45\textwidth]{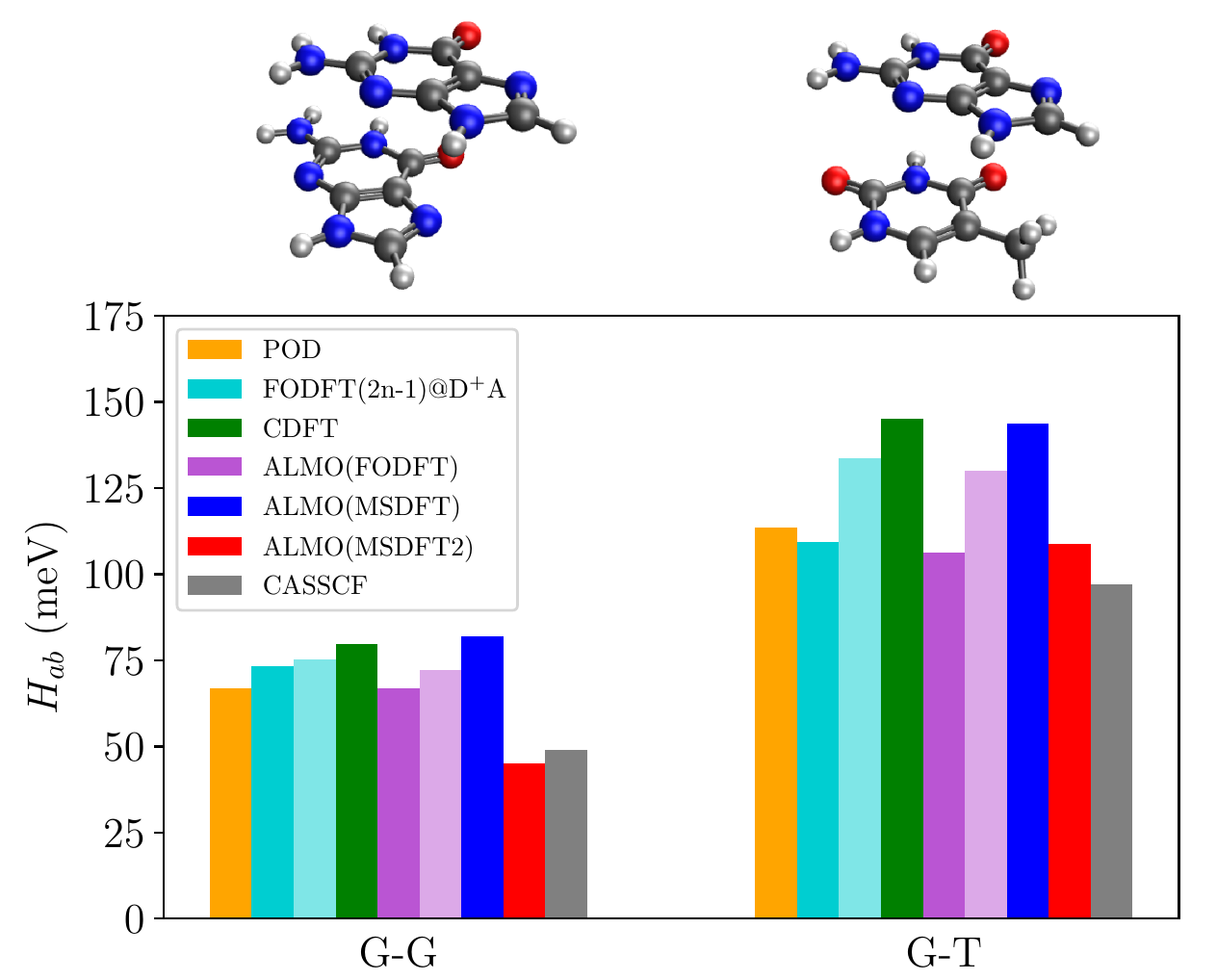}
    \caption{Performance of different methods for hole transfer in asymmetric $\pi$-stacked complexes form by DNA bases: guanine-guanine (G-G) and guanine-thymine (G-T). The DFT calculations are performed at the LRC-$\omega$PBEh/6-31+G(d) level of theory. For FODFT($2n\mathrm{-}1$)@$D^+A$ and ALMO(FODFT), the results with the first guanine fragment as the donor (positively charged) are shown in the darker color, and the results with the charge population reversed are shown in lighter color.}
    \label{fig:dna}
\end{figure}

The systems contained in the HAB11 HT and HAB7- ET test sets are all symmetric systems allowing Eq.~\ref{eq:half_gap} to be used to extract their diabatic couplings from excitation energies calculated by high-level electronic structure methods. However, it is important to also assess how these methods fare when faced with complexes with no point-group symmetry. Hence, as the last example, we investigate how these diabatization methods perform in evaluating the diabatic couplings for HT in two $\pi$-stacked complexes formed by DNA nucleobases: guanine-guanine (G-G) and guanine-thymine (G-T), using the geometries provided in Ref.~\citenum{Blancafort2006}. These systems also serve as models for investigating the mechanism of DNA-mediated charge transport, which has important implications in biochemical processes such as oxidative DNA damage and repair.\cite{ONeill2004} 

Figure~\ref{fig:dna} shows the resulting diabatic couplings for these two systems. We compare our values to the previous benchmark calculations performed using CASSCF(11,12)/6-31G(d) followed by a GMH diabatization.\cite{Blancafort2006} For the G-G complex, which breaks symmetry due to the imperfect $\pi$-stacking of the two guanine monomers, ALMO(MSDFT2) yields a diabatic coupling with an error of only 8\% relative to the CASSCF reference with the next best method ALMO(FODFT) giving an error of 37\% and the original ALMO(MSDFT) giving the largest error of 67\%. As discussed in Sec.~\ref{sec:method}, the results obtained from FODFT-based methods depend on the initial charge-localized reference state. For asymmetric systems, such as the guanine pair considered here, FODFT-based methods therefore produce different results when choosing the reference state as G$^+$-G or G-G$^+$. The two values obtained for the FODFT($2n\mathrm{-}1$)@$D^+A$ and ALMO(FODFT) methods are shown in Fig.~\ref{fig:dna} as the different shades of the cyan and purple bars, respectively. Since in the G-G system the symmetry is broken only by the arrangement of the monomers, the differences in the coupling obtained are relatively small: 2.1 meV for FODFT($2n\mathrm{-}1$)@$D^+A$ and 5.2 meV for ALMO(FODFT). However, when one considers the explicitly asymmetric G-T complex, the gap between the results using the two different reference states (G$^+$-T and G-T$^+$) increases substantially to 24.5 meV for FODFT($2n\mathrm{-}1$)@$D^+A$ and 23.7 meV for ALMO(FODFT), which is a change of 22\% in the result obtained depending on the reference chosen. ALMO(MSDFT2), does not suffer from this issue since it encodes the symmetry in the construction of the diabatic Hamiltonian (Eq.~\ref{eq:msdft2}) and yields an error of 12\% relative to the CASSCF result.

\section{Conclusions}

In this work, we have shown that the ALMO(MSDFT2) and ALMO(FODFT) approaches introduced here possess a number of advantages over other DFT-based diabatization methods (POD, FODFT, and CDFT). In particular, by benchmarking on the HAB11 HT and HAB7- ET datasets, we showed that ALMO(MSDFT2) yields the best accuracy among all investigated approaches, with the smallest MUREs of $< 5\%$ when paired with RSH functionals. Indeed, with such small errors, the ALMO(MSDFT2) approach comes within the error bars of the high-level multireference methods traditionally used to provide the benchmarks for these systems. In addition, owing to its more internally consistent treatment of the XC contribution to the diabatic coupling compared to the original MSDFT scheme, the ALMO(MSDFT2) method is able to give accurate diabatic couplings even when combined with lower-tier XC functionals with GGA MUREs only rising by 3--6\% and shows systematic convergence with respect to the basis set employed. Using DNA base pairs as an example, we have further demonstrated that the advantage of the symmetry encoded in the construction of the ALMO(MSDFT2) diabatic Hamiltonian allows it to unambiguously and accurately treat asymmetric charge transfer. Finally, since the diabatic states in the ALMO approach are variationally optimized at the full system level, they allow for the associated forces to be readily computed as well as capture the energetic stabilization arising from the polarization of the donor and acceptor species in each other's presence.

The ALMO(MSDFT2) method therefore should provide a useful tool for constructing ab initio diabatic potential energy surfaces in large condensed phase environments, where only the lowest tiers of the DFT hierarchy are affordable, facilitating the simulation of nonadiabatic processes in these systems.

\section*{Supplementary Material}
Full statistical errors for the HAB11 and HAB7- datasets; additional benchmark results for the exponential decay rate ($\beta$), basis set dependence, and the performance of TDDFT and SF-TDDFT; results for the electron and hole transfer in the pentacene dimer.

\begin{acknowledgments}
This material is based upon work supported by the National Science Foundation under Grant No.~CHE-1652960. T.E.M also acknowledges support from the Camille Dreyfus Teacher-Scholar Awards Program.
\end{acknowledgments}

%


\clearpage

\setcounter{section}{0}
\setcounter{equation}{0}
\setcounter{figure}{0}
\setcounter{table}{0}
\setcounter{page}{1}

\renewcommand{\theequation}{S\arabic{equation}}
\renewcommand{\thefigure}{S\arabic{figure}}
\renewcommand{\thetable}{S\arabic{table}}
\renewcommand{\thepage}{S\arabic{page}}
\renewcommand{\bibnumfmt}[1]{$^{\mathrm{S#1}}$}
\renewcommand{\citenumfont}[1]{S#1}

\title{Supporting Information for ``Accurate and efficient DFT-based diabatization for hole and electron transfer using absolutely localized molecular orbitals" }

{\maketitle}

\onecolumngrid

\section{Additional benchmark results}

\begin{figure*}[h!]
    \centering
    \includegraphics[width=0.85\textwidth]{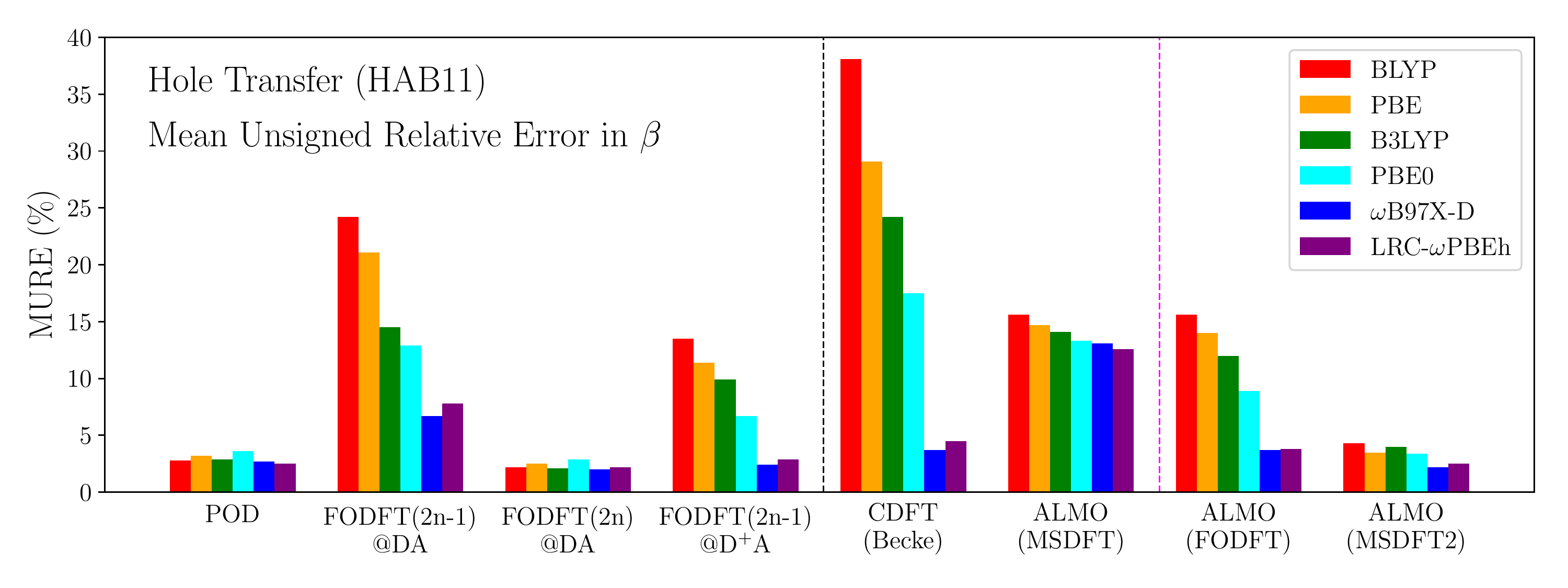}
    \caption{Performance of the diabatization schemes as indicated by their MUREs for the exponential decay constant ($\beta$) in the HAB11 hole transfer dataset. Methods on the right of the black dashed line are ones that can produce variationally optimized diabatic states, and those on the right of the magenta dashed line are the ones proposed in this work.}
    \label{fig:hab11_beta}
\end{figure*}

\begin{figure*}[h!]
    \centering
    \includegraphics[width=0.85\textwidth]{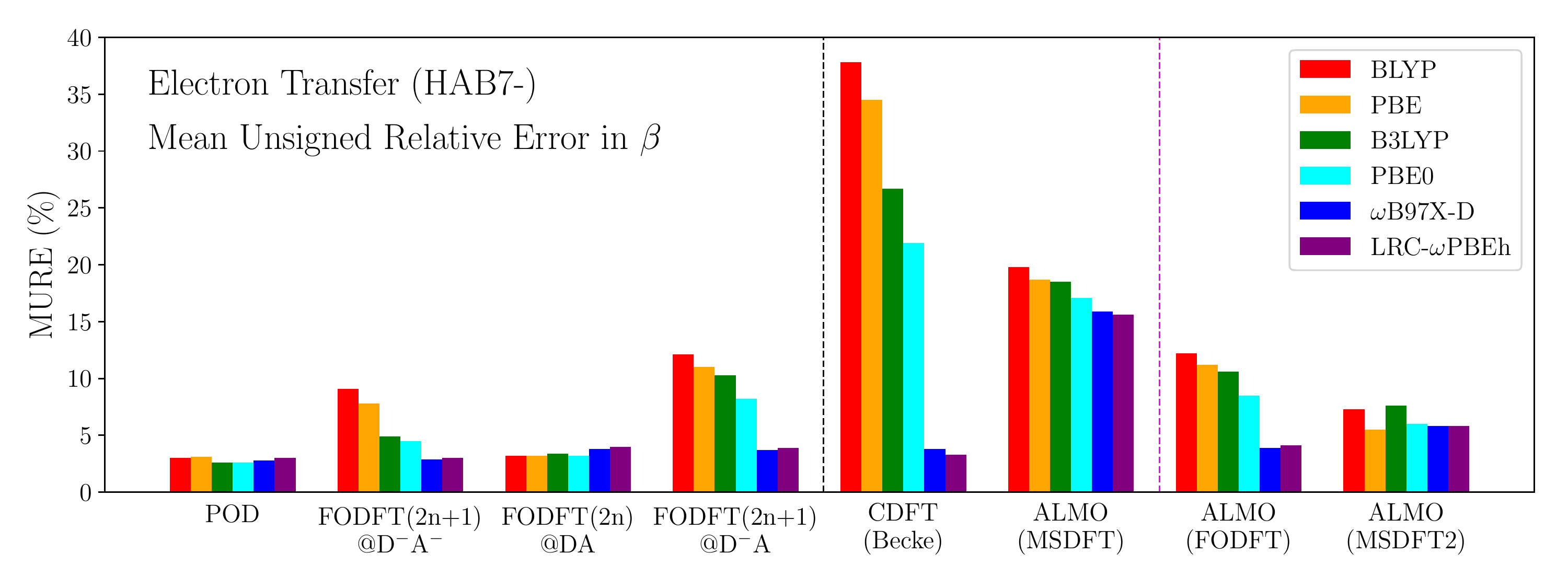}
    \caption{Performance of the diabatization schemes as indicated by their MUREs for the exponential decay constant ($\beta$) in the HAB7- electron transfer dataset. Methods on the right of the black dashed line are ones that can produce variationally optimized diabatic states, and those on the right of the magenta dashed line are ones proposed in this work.}
    \label{fig:hab7_beta}
\end{figure*}

\begin{figure}[h!]
    \centering
    \includegraphics[width=0.5\textwidth]{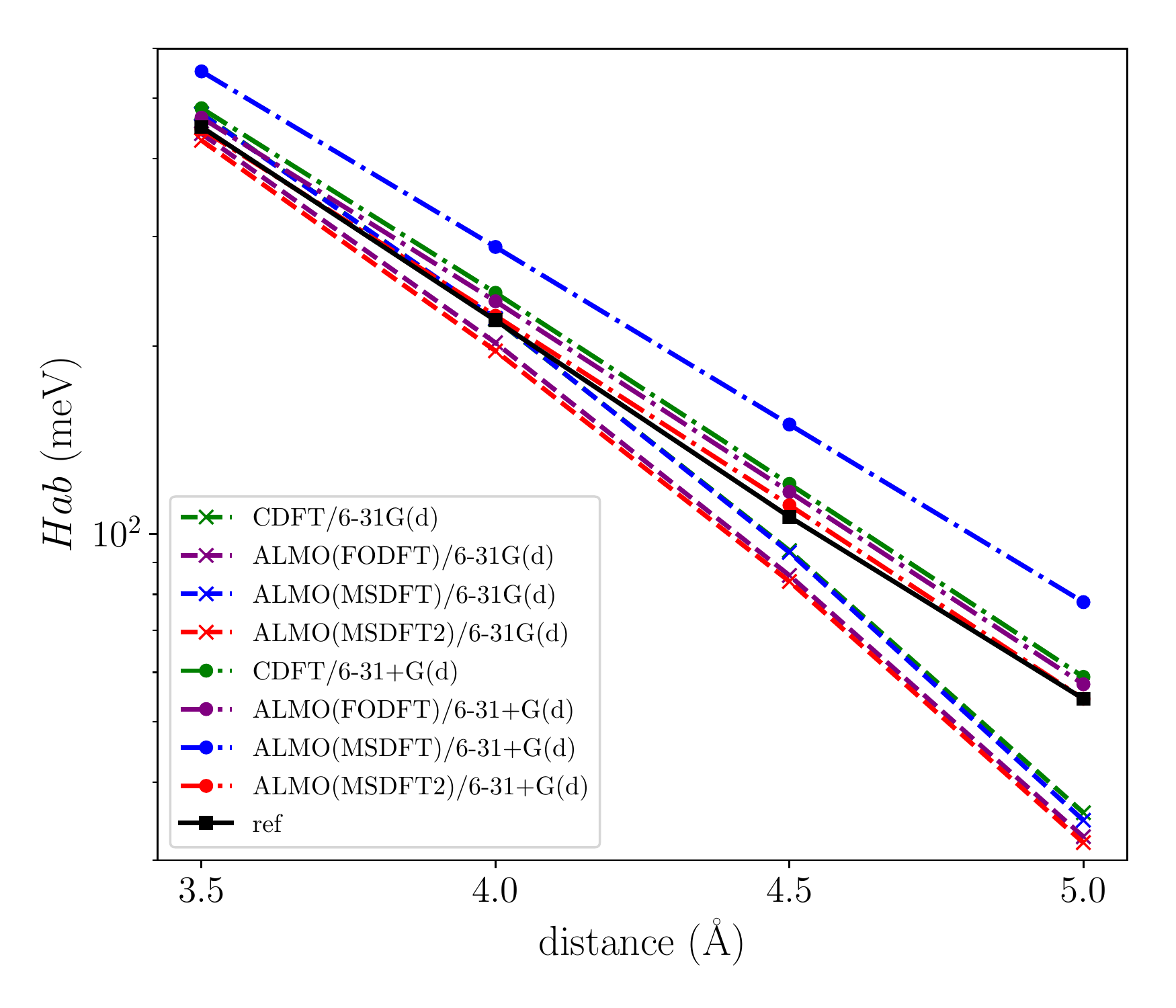}
    \caption{Demonstration of the inability of the 6-31G(d) basis set to capture the correct long-range decay behavior of $\vert H_{ab} \vert$. The calculations were performed with the LRC-$\omega$PBEh functional.}
\end{figure}

\begin{figure}[t!]
    \centering
    \includegraphics[width=0.5\textwidth]{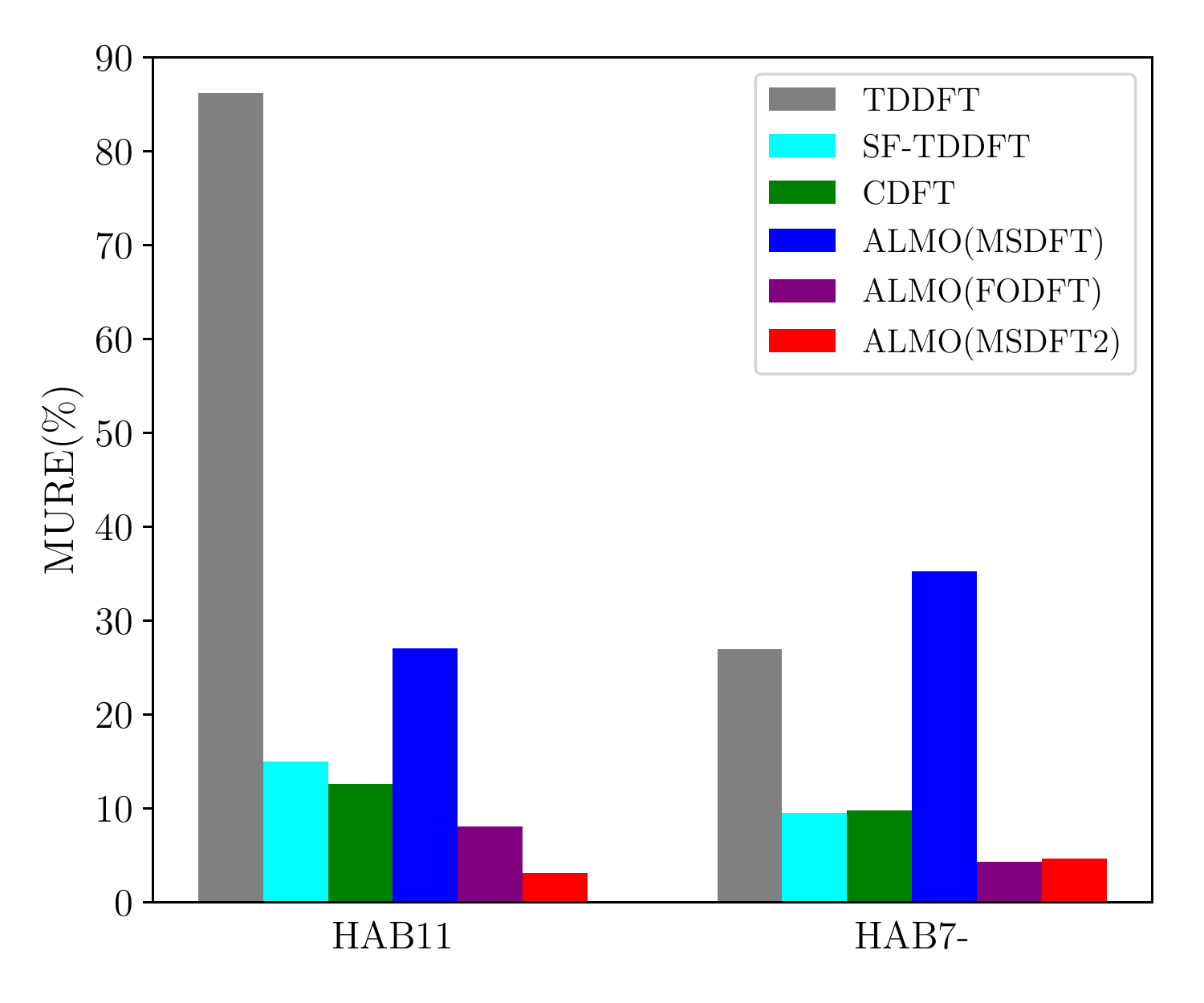}
    \caption{Comparison of the MUREs of TDDFT and SF-TDDFT against those of DFT-based diabatization schemes that directly construct variationally optimized diabatic states [CDFT, ALMO(MSDFT), ALMO(FODFT), and ALMO(MSDFT2)]. All calculations are performed with the $\omega$B97X-D functional and the 6-31+G(d) basis.}
\end{figure}

\begin{figure*}[t!]
    \centering
    \begin{tabular}{cc}
    \includegraphics[width=0.45\textwidth]{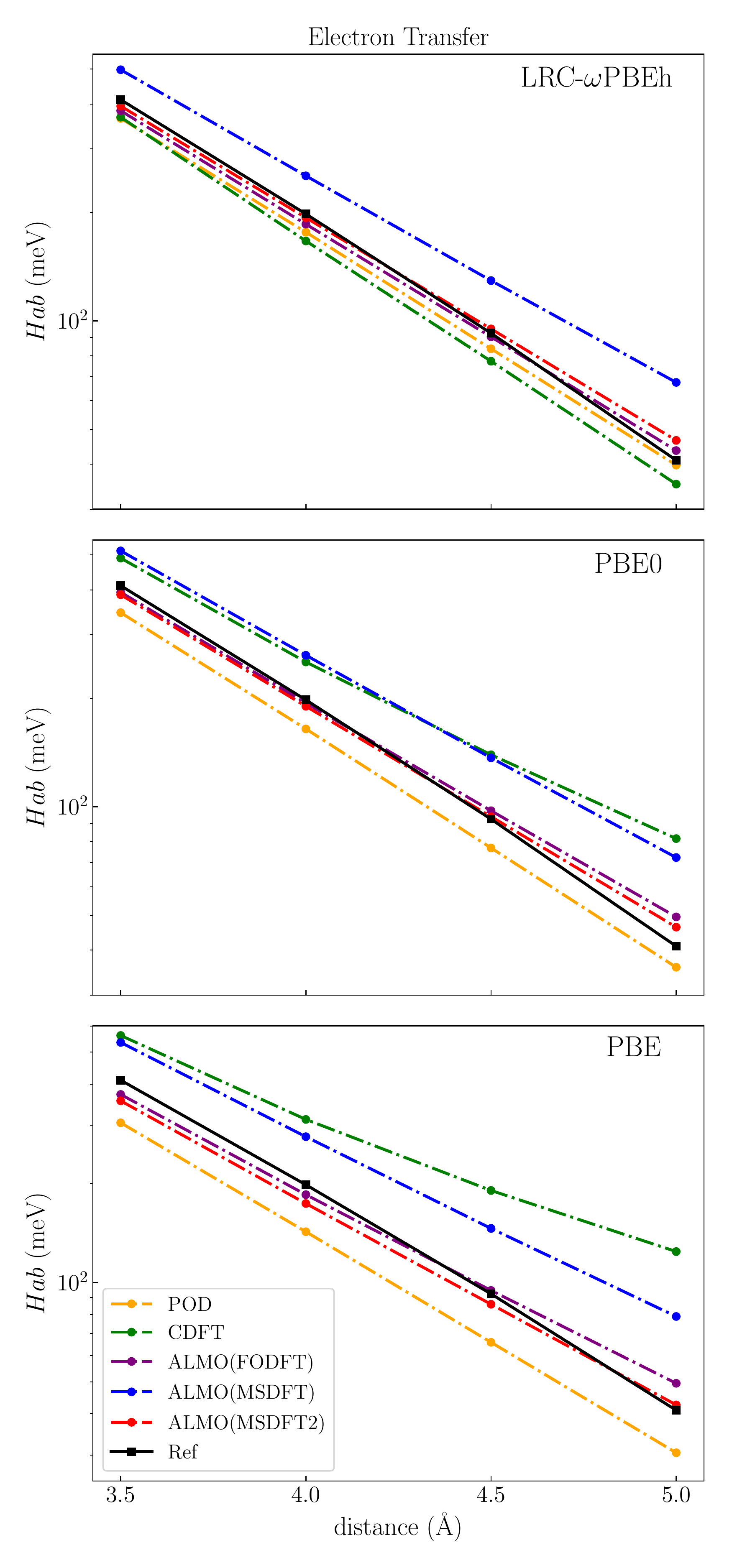} &
    \includegraphics[width=0.45\textwidth]{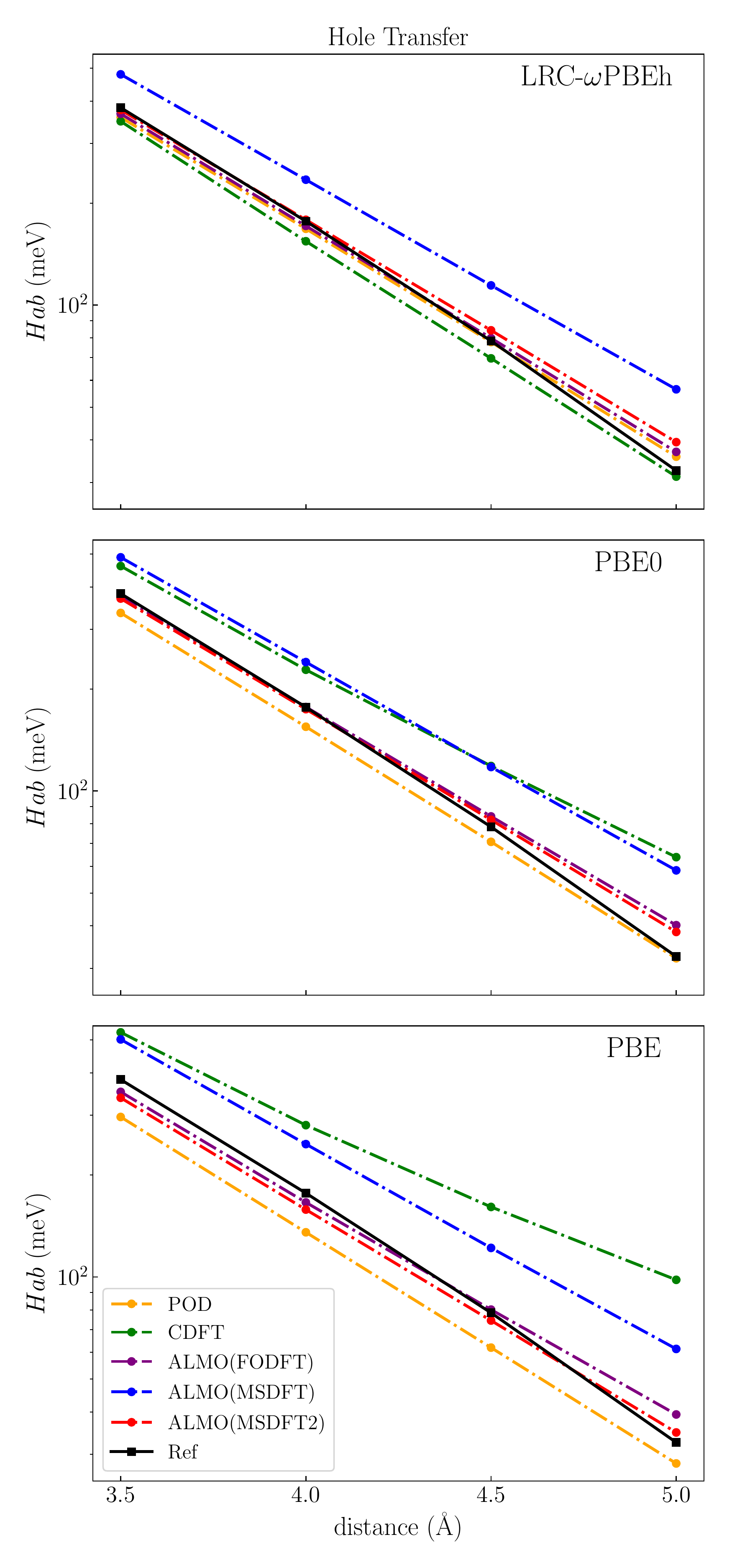}
    \end{tabular}
    \caption{Effect of the employed XC functional on the performance of DFT-based diabatization schemes in capturing the distance dependence of $\vert H_{ab} \vert$ for the ET (left panels) and HT (right panels) in the pentacene dimer. The y-axis ($\vert H_{ab} \vert$) is in a logarithmic scale. Note that the reference values, which were taken from Ref.~56, might be too small at the longer distances as the $\vert H_{ab} \vert$ values notably deviate from the expected exponential decay behavior.}
    \label{fig:pentacene_ET_HT}
\end{figure*}

\begin{table}[t!]
    \centering
    \caption{Mean signed error (MSE), mean unsigned error (MUE), mean signed relative error (MSRE), and mean unsigned relative error (MURE) of the DFT-based diabatization schemes for the HAB11 hole transfer dataset evaluated with six different density functionals and the 6-31+G(d) basis set. The absolute errors are in meV. }
    \begin{tabularx}{\textwidth}{c *{8}{Y}}
    \hline
     &  & BLYP & PBE & B3LYP & PBE0 & $\omega$B97X-D & LRC-$\omega$PBEh\\
     \hline
    POD & MSE & -56.17 & -57.42 & -41.54 & -39.68 & -19.16 & -23.96\\
    & MUE & 56.17 & 57.42 & 41.54 & 39.68 & 19.16 & 23.96\\
    & MSRE & -27.3\% & -28.3\% & -20.5\% & -20.3\% & -9.6\% & -11.8\% \\
    & MURE & 27.3\% & 28.3\% & 20.5\% & 20.3\% & 9.6\% & 11.8\% \\
    \hline
    FODFT($2n\mathrm{-}1$)@$DA$ & MSE & -69.29 & -69.29 & -50.26 & -47.38 & -27.16 & -37.95\\
    & MUE & 69.29 & 69.29 & 50.26 & 47.38 & 27.16 & 37.95\\
    & MSRE & -39.9\% & -39.2\% & -28.9\% & -27.2\% & -15.7\% & -21.1\% \\
    & MURE & 39.9\% & 39.2\% & 28.9\% & 27.2\% & 15.7\% & 21.1\% \\
    \hline
    FODFT($2n$)@$DA$ & MSE & -47.97 & -49.02 & -34.17 & -32.78 & -14.20 & -19.06\\
    & MUE & 47.97 & 49.02 & 34.17 & 32.78 & 14.42 & 19.15\\
    & MSRE & -22.8\% & -23.6\% & -16.1\% & -16.2\% & -6.1\% & -8.3\% \\
    & MURE & 22.8\% & 23.6\% & 16.1\% & 16.2\% & 6.6\% & 8.5\%
    \\
    \hline
    FODFT($2n\mathrm{-}1$)@$D^+A$ & MSE & 14.48 & 7.52 & 12.24 & 4.62 & 3.31 & -5.66\\
    & MUE & 17.94 & 13.35 & 13.87 & 7.95 & 6.26 & 6.86\\
     & MSRE & 14.7\% & 9.8\% & 11.4\% & 5.7\% & 2.4\% & -1.5\% \\
    & MURE & 15.7\% & 11.6\% & 11.9\% & 6.7\% & 3.8\% & 3.5\% \\
    \hline
    CDFT(Becke) & MSE & 172.74 & 159.80 & 105.37 & 72.57 & 22.15 & 12.79\\
    & MUE & 177.68 & 159.80 & 105.37 & 72.57 & 22.15 & 12.79\\
    & MSRE & 126.9\% & 109.6\% & 72.2\% & 48.3\% & 12.6\% & 8.5\% \\
    & MURE & 128.1\% & 109.6\% & 72.2\% & 48.3\% & 12.6\% & 8.5\% \\
    \hline
    ALMO(MSDFT) & MSE & 43.30 & 40.77 & 41.61 & 35.76 & 37.27 & 35.21\\
    & MUE & 43.30 & 40.77 & 41.61 & 35.76 & 37.27 & 35.21\\
    & MSRE & 32.0\% & 30.1\% & 30.1\% & 26.4\% & 27.1\% & 25.6\% \\
    & MURE & 32.0\% & 30.1\% & 30.1\% & 26.4\% & 27.1\% & 25.6\% \\
    \hline
    ALMO(FODFT) & MSE & 33.97 & 27.37 & 26.89 & 17.36 & 12.71 & 1.72\\
    & MUE & 34.81 & 28.64 & 27.14 & 17.71 & 12.90 & 7.46\\
    & MSRE & 26.2\% & 21.6\% & 20.2\% & 13.4\% & 8.0\% & 2.7\% \\
    & MURE & 26.4\% & 22.0\% & 20.2\% & 13.5\% & 8.1\% & 4.8\% \\
    \hline
    ALMO(MSDFT2) & MSE & -13.79 & -10.99 & -1.28 & 0.82 & -1.60 & 1.44\\
    & MUE & 14.51 & 11.80 & 10.57 & 5.93 & 5.32 & 5.33\\
    & MSRE & -4.5\% & -3.7\% & 1.5\% & 2.2\% & 0.3\% & 2.0\%\\
    & MURE & 5.8\% & 5.1\% & 5.5\% & 4.1\% & 3.1\% & 3.5\% \\
    \hline
    \end{tabularx}
    \label{tab:hab11_errors}
\end{table}

\begin{table}[t!]
    \centering
    \caption{MSE, MUE, MSRE, and MURE of the DFT-based diabatization schemes for the HAB7- electron transfer dataset evaluated with six different density functionals and the 6-31+G(d) basis set. The absolute errors are in meV. }
    \begin{tabularx}{\textwidth}{c *{8}{Y}}
    \hline
    &  & BLYP & PBE & B3LYP & PBE0 & $\omega$B97X-D & LRC-$\omega$PBEh\\
    \hline
    POD & MSE & -49.96 & -50.65 & -33.70 & -30.88 & -16.25 & -21.22\\
    & MUE & 49.96 & 50.65 & 33.70 & 30.88 & 16.25 & 21.22\\
    & MSRE & -28.9\% & -29.4\% & -18.9\% & -17.5\% & -8.3\% & -1.1\% \\
    & MURE & 28.9\% & 29.4\% & 18.9\% & 17.5\% & 8.3\% & 1.1\% \\
    \hline
    FODFT($2n\mathrm{+}1$)@$D^-A^-$ & MSE & -47.96 & -46.99 & -30.84 & -27.19 & -11.28 & -16.28\\
    & MUE & 47.96 & 46.99 & 30.84 & 27.19 & 11.53 & 16.37\\
    & MSRE & -29.7\% & -28.8\% & -18.5\% & -16.4\% & -5.4\% & -8.4\% \\
    & MURE & 29.7\% & 28.8\% & 18.5\% & 16.4\% & 6.1\% & 8.7\% \\
    \hline
    FODFT($2n$)@$DA$ & MSE & -36.88 & -37.75 & -19.54 & -17.11 & -1.82 & -7.59\\
    & MUE & 36.88 & 37.75 & 19.55 & 17.16 & 4.45 & 8.76\\
    & MSRE & -21.0\% & -21.5\% & -10.0\% & -8.9\% & 1.0\% & -2.1\% \\
    & MURE & 21.0\% & 21.5\% & 10.0\% & 9.1\% & 3.7\% & 5.0\% \\
    \hline
    FODFT($2n\mathrm{+}1$)@D$^-$A & MSE & -13.90 & -15.44 & -5.24 & -5.65 & -7.00 & -11.73\\
    & MUE & 18.42 & 18.73 & 11.26 & 9.68 & 7.86 & 12.29\\
    & MSRE & -1.5\% & -3.1\% & 2.8\% & 1.3\% & -2.0\% & -4.6\% \\
    & MURE & 10.2\% & 9.8\% & 8.2\% & 6.7\% & 4.3\% & 6.2\% \\
    \hline
    CDFT(Becke) & MSE & 122.53 & 115.00 & 74.49 & 57.51 & -14.58 & -19.49\\
    & MUE & 122.53 & 115.00 & 74.49 & 57.51 & 14.71 & 19.63\\
    & MSRE & 116.1\% & 103.8\% & 66.6\% & 50.7\% & -9.3\% & -11.6\% \\
    & MURE & 116.1\% & 103.8\% & 66.6\% & 50.7\% & 9.8\% & 12.1\% \\
    \hline
    ALMO(MSDFT) & MSE & 57.14 & 53.65 & 49.82 & 44.43 & 41.05 & 39.12\\
    & MUE & 57.14 & 53.65 & 49.82 & 44.43 & 41.05 & 39.12\\
    & MSRE & 48.5\% & 45.4\% & 42.8\% & 38.3\% & 35.3\% & 33.8\% \\
    & MURE & 48.5\% & 45.4\% & 42.8\% & 38.3\% & 35.3\% & 33.8\% \\
    \hline
    ALMO(FODFT) & MSE & -9.98 & -11.19 & -2.08 & -2.67 & -5.89 & -10.63\\
    & MUE & 15.98 & 15.86 & 10.01 & 8.26 & 7.23 & 11.47\\
    & MSRE & 1.0\% & -0.4\% & 4.9\% & 3.3\% & -1.1\% & -3.7\% \\
    & MURE & 10.0\% & 9.4\% & 8.7\% & 7.0\% & 4.3\% & 6.0\% \\
    \hline
    ALMO(MSDFT2) & MSE & -23.55 & -23.12 & -10.49 & -7.00 & -3.72 & -3.34\\
    & MUE & 24.18 & 23.43 & 12.89 & 9.34 & 6.73 & 6.36\\
    & MSRE & -9.7\% & -10.3\% & -1.8\% & -0.7\% & 1.3\% & 1.4\% \\
    & MURE & 11.1\% & 11.0\% & 6.8\% & 5.6\% & 4.7\% & 4.7\% \\
    \hline
    \end{tabularx}
    \label{tab:hab7_errors}
\end{table}
	
\end{document}